\documentclass[fleqn,10pt]{wlscirep}
\usepackage[T1]{fontenc}

\PassOptionsToPackage{utf8}{inputenc}

\PassOptionsToPackage{bookmarks,colorlinks}{hyperref}

\usepackage[T1]{fontenc}

\usepackage{amsthm}

\usepackage{amssymb,amsthm,bm,mathtools,steinmetz}

\usepackage{graphicx}
\usepackage{pgfplots}
\pgfplotsset{compat=1.18}
\usetikzlibrary{shapes.multipart,intersections}

\usepackage{booktabs}
\usepackage{float}

\usepackage{mathrsfs}
\usepackage{textcomp}
\usepackage{upgreek}
\usepackage{xspace}
\usepackage{enumitem}
\usepackage{algpseudocode}
\usepackage{balance}
\usepackage{verbatim}
\usepackage{url}

\title{Frozen differential scattering in reconfigurable complex media}

\author[1,*]{Philipp del Hougne}
\affil[1]{ Univ Rennes, CNRS, IETR - UMR 6164, F-35000, Rennes, France}

\affil[*]{philipp.del-hougne@univ-rennes.fr}

\begin{abstract}
The sensitivity of transmission to the input wavefront is a hallmark feature of complex media and the basis for wavefront shaping techniques. Yet, intriguing special cases exist in which the output wavefront is ``frozen'' (agnostic to the input wavefront). This happens when special structure in the complex medium collapses the rank of its transmission matrix to unity. Here, we unveil that an analogous phenomenon exists more universally for differential scattering (including reflection) in reconfigurable complex media. Specifically, for a localized perturbation, the differential scattering matrix of any complex medium has rank one. One consequence is that the differential output signal is perfectly coherent irrespective of the input wavefront's coherence. Moreover, the thermal noise emitted into the frozen differential output mode has a particular structure that can be exploited for thermal noise management. We experimentally evidence frozen differential scattering in a rich-scattering wireless link parametrized by a programmable meta-atom. Then, we demonstrate ``customized freezing'' by optimizing the configuration of additional programmable meta-atoms that parametrize the wireless link, as envisioned for 6G networks. We impose particular shapes of the frozen differential output mode, and maximize its signal-to-thermal-noise ratio. Potential applications include filtering and stabilization of differential wavefronts, as well as imaging, sensing, and communication in complex media.
\end{abstract}
\begin{document}

\flushbottom
\maketitle

\thispagestyle{empty}

\section{Introduction}

\begin{figure*}
\centering
\includegraphics[width=\linewidth]{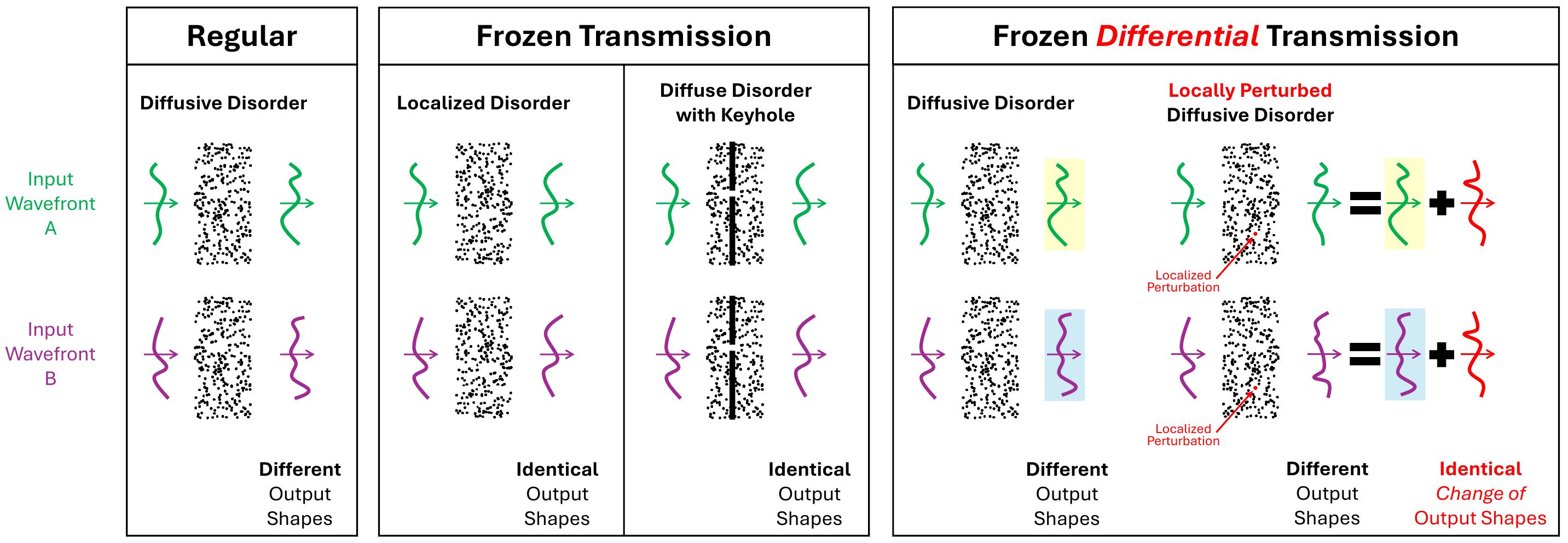}
\caption{Schematic of wavefront freezing phenomena. Left: conventional full-rank transmission where different inputs yield different outputs. Middle: frozen transmission in special static media (Anderson-localized or keyhole), where the output is ``frozen'' in the same shape no matter the input. Right: locally perturbed complex medium, where the \textit{change in} output is ``frozen'' no matter the input. This illustration of differential freezing is limited to transmission for simplicity, but differential freezing applies to reflection as well. }
\label{Fig1}
\end{figure*}

The propagation of waves in complex media is subject to immensely complex, seemingly random, multiple scattering. In the optical regime, strongly scattering biological tissues, turbid media like clouds, and disordered waveguides are prominent examples of complex media; in the microwave regime, scattering radio environments are a relevant example for wireless communications. An impinging wavefront is completely scrambled as a result of its interaction with a complex medium. This scrambling was long seen as a severe obstacle for imaging and communications, across the electromagnetic spectrum. However, \textit{the scrambled output wavefront is highly sensitive to the impinging input wavefront}, as illustrated in the left panel in Fig.~\ref{Fig1}. The mapping from the latter to the former is deterministic and linear (within the cases discussed in this paper). Therefore, coherent control of the input wavefront, a technique known as wavefront shaping, enables us to tailor the complicated wave interferences inside complex media. As a result, the seemingly detrimental effect of random scattering can be compensated, and even leveraged to outperform the wave control possible in perfectly homogeneous media. Over the past decades, wavefront shaping has emerged as a pivotal tool for MIMO wireless communications~\cite{moustakas2000communication} and controlling light transport in complex media~\cite{cao2022shaping}.

Yet, the sensitivity of the output wavefront to the input wavefront on which wavefront shaping relies is \textit{not} universal in complex media. Indeed, the output wavefront's shape is ``frozen'' (i.e., agnostic to the input wavefront) in certain complex media, as illustrated in the middle panel of Fig.~\ref{Fig1}. A striking explicit observation of this phenomenon is the invariance of the speckle pattern at the output side of a random waveguide in the regime of Anderson localization~\cite{shi2012transmission}. It turns out that this phenomenon is associated with a special mathematical structure of the transmission matrix (TM) that maps input wavefront to output wavefront. Specifically, there is only one dominant transmission eigenvalue in the regime of Anderson localization, implying that the rank of the TM is approximately unity~\cite{shi2012transmission,davy2012focusing,wang2011transport,pena2014single,leseur2014probing}. This peculiar TM property has also been predicted~\cite{chizhik2000capacities,chizhik2002keyholes} and experimentally observed~\cite{almers2003measurement,almers2006keyhole} in so-called keyhole MIMO wireless links, although it has to date not been linked to frozen wavefronts in these cases. More recently, there have been efforts to deliberately engineer specific periodic waveguides and gratings with rank-one TMs that yield frozen wavefronts~\cite{salemeh2025invariance,salemeh2025freezing}.

In this paper, we unveil an analogous phenomenon of input-wavefront invariance in \textit{reconfigurable} complex media that we illustrate in the right panel in Fig.~\ref{Fig1}. The phenomenon we describe generically applies to scattering (transmission and/or reflection) in any complex medium across broad frequency ranges. This is in contrast to the aforementioned examples of frozen wavefronts in transmission (\textit{but not reflection}) in \textit{special}, \textit{static} complex media, some of which are limited to \textit{narrow} bands. 
The phenomenon on which we report originates from the fact that the change of the scattering matrix due to a localized perturbation is of rank one for any complex medium. As a result, the change of the output wavefront, referred to as the \textit{differential} output wavefront, is frozen. As a prominent, contemporary example, we consider a smart 6G radio environment wherein a programmable metasurface parametrizes a wireless MIMO link. The meta-atoms' programmability relies on PIN diodes, which are tunable \textit{lumped} (i.e., point-like) components that are accurately described as localized perturbers within a rigorous multiport-network system model. If scattering is predominantly in the forward direction, as in many reconfigurable optical systems like multi-plane light converters (MPLCs)~\cite{morizur2010programmable,berkhout2010efficient,zhang2023multi} and diffractive neural networks~\cite{zhou2021large}, the perturbation can be of the size of a resolution cell, whose size can largely exceed the wavelength. Incidentally, large MPLCs have been shown to behave like random scattering media~\cite{boucher2021full}.

Differential wave-scattering measurements underpin a wide spectrum of practices that isolate weak changes while suppressing static clutter. In optics, differential light scattering enables label-free identification of microorganisms~\cite{wyatt1968differential,wyatt1969identification}; chiroptical circular intensity differential scattering and related approaches probe handedness-selective responses~\cite{bustamante1980circular,bustamante1980circularII,bustamante1983circular}; and differential reflectance spectroscopy sensitively characterizes thin films and surfaces~\cite{forker2012optical}. In remote sensing, coherent change detection radar (including polarimetric variants) exploits complex-valued differences between passes to flag localized scene changes~\cite{7163589,wahl2016new,8401705}. In communications, differential detection (e.g., differential phase-shift keying) likewise extracts information from state differences rather than absolute phase~\cite{alouini_dpsk}. 
Our findings in this paper show that, for a sufficiently localized perturbation, all such differential wave-scattering measurements inherit a universal rank-one structure and the differential signal always lies in a single ``frozen'' spatial mode. This insight can explain robustness to unknown inputs, gives a principled reason why differential schemes concentrate information into a single mode, and can be directly exploited for conceiving an optimal differential read-out strategy.

The remainder of this paper is organized as follows. We begin by explaining the principle of frozen differential scattering. Then, we report an experimental observation thereof. We go on to examine its influence on coherence and thermal noise emission. Finally, we experimentally demonstrate the customization of frozen differential scattering to impose particular shapes of the frozen differential output mode, or to maximize the signal-to-thermal-noise ratio of freezing-aware differential measurements.

\section{Principle of wavefront invariance}
\label{sec_theory}

\subsection{Frozen transmission in static complex media}
\label{subsec_static_principle}

The interaction of waves with a linear, static, passive complex medium is characterized by the medium's scattering matrix $\mathbf{S}\in\mathbb{C}^{N_\mathrm{A} \times N_\mathrm{A}}$ that maps the incoming wavefront $\mathbf{x}\in\mathbb{C}^{N_\mathrm{A}}$ to the outgoing wavefront $\mathbf{y}\in\mathbb{C}^{N_\mathrm{A}}$: $\mathbf{y}=\mathbf{S}\mathbf{x}$. Each of the $N_\mathrm{A}$ wavefront entries is associated with a distinct channel that may be localized in space (e.g., waveguide channels) or in momentum space (e.g., angular momentum channels).
If waves are only injected via $N_\mathrm{T}$ channels, and only detected via the remaining $N_\mathrm{R}=N_\mathrm{A}-N_\mathrm{T}$ channels, then the observable wave-matter interactions are characterized by the TM $\mathbf{T}\in\mathbb{C}^{N_\mathrm{R}\times N_\mathrm{T}}$ which is an off-diagonal block of $\mathbf{S}$: $\mathbf{T}=\mathbf{S}_\mathcal{RT}$, and $\mathbf{y}_\mathcal{R} = \mathbf{T}\mathbf{x}_\mathcal{T}$. We here use a compact notation, defined in the Methods, wherein $\mathcal{T}$ and $\mathcal{R}$ are the sets of channel indices associated with injection and detection channels, respectively.

The known examples of frozen wavefronts discussed in the introduction all relate to static complex media whose TMs have a common, special mathematical structure. To understand this structure, a singular value decomposition (SVD) of the TM is required. The SVD allows us to express $\mathbf{T}$ as the product of three matrices: $\mathbf{T}=\mathbf{U}\mathbf{\Sigma}\mathbf{V}^\dagger$, where $\mathbf{\Sigma}=\mathrm{diag}([\tau_1, \tau_2, \dots, \tau_{\tilde{N}}])$ is a diagonal matrix containing the singular values of $\mathbf{T}$ in descending order and $\tilde{N}=\mathrm{min}(N_\mathrm{R},N_\mathrm{T})$. The $i$th singular value $\tau_i$ is associated with a left singular vector $\mathbf{u}_i$ (the $i$th column of $\mathbf{U}$) and a right singular vector $\mathbf{v}_i$ (the $i$th column of $\mathbf{V}$). These singular vectors define orthogonal input/output modes of the complex medium~\cite{miller2019waves}.
The abstract mathematical structure of the TM shared by the various classes of static complex media described in the introduction is that their first singular value dominates: $\tau_1 \gg \tau_2$. Consequently, their TM is approximately of rank one: $\mathbf{T}\approx \mathbf{u}_1 \tau_1 \mathbf{v}_1^\dagger$. The transmitted wavefront is thus $\mathbf{y}_\mathcal{R} \approx \mathbf{u}_1 \tau_1 \mathbf{v}_1^\dagger \mathbf{x}_\mathcal{T} = \alpha(\mathbf{x}_\mathcal{T}) \mathbf{u}_1$, where $\alpha(\mathbf{x}_\mathcal{T})=\tau_1 \mathbf{v}_1^\dagger \mathbf{x}_\mathcal{T}$ is a complex-valued scalar. It follows by inspection that the transmitted wavefront $\mathbf{y}_\mathcal{R}$ is always approximately collinear with $\mathbf{u}_1$, and the choice of input wavefront $\mathbf{x}_\mathcal{T}$ only affects a complex-valued scaling factor. This is precisely why the shape of $\mathbf{y}_\mathcal{R}$ is frozen, i.e., agnostic to $\mathbf{x}_\mathcal{T}$.

To rigorously quantify the extent to which one singular value dominates the TM, we compute the TM's effective rank $R(\mathbf{T}) = \mathrm{exp} \left(-\sum_{k=1}^{\tilde{N}} \hat{\tau}_k \mathrm{ln}(\hat{\tau}_k)\right)$~\cite{roy2007effective}, where $\hat{\tau}_k = \tau_k / \sum_{l=1}^{\tilde{N}} {\tau_l} $, and the TM's participation number $P(\mathbf{T})=\left(\sum_{k=1}^{\tilde{N}}{ \tau_k^2 }\right)^2/\sum_{k=1}^{\tilde{N}}{ \tau_k^4 }$~\cite{davy2012focusing}.\footnote{The statistical ratio defined in Ref.~\cite{pena2014single} to detect the regime of Anderson localization is specific to random media because it relies on assumptions about the statistical properties of $\mathbf{T}$.} Note that $1 \leq R \leq \tilde{N}$ and $1\leq P \leq \tilde{N}$; both $R$ and $P$ tend to unity when one singular value dominates, indicating that the transmitted wavefront is frozen. A more direct metric related to frozen wavefronts is the average collinearity of two output wavefronts $\mathbf{y}_{\mathcal{R}}$ and $\mathbf{\stackrel{\circ}{y}}_{\mathcal{R}}$ associated with different input wavefronts: $C(\mathbf{T}) = \langle | \mathbf{y}_{\mathcal{R}}^\dagger \mathbf{\stackrel{\circ}{y}}_{\mathcal{R}} | / \lVert \mathbf{y}_{\mathcal{R}} \rVert \lVert \mathbf{\stackrel{\circ}{y}}_{\mathcal{R}} \rVert  \rangle$~\cite{leseur2014probing}, where the averaging is over different random choices of normalized input wavefronts. Note that $0\leq C \leq 1$; $C$ tends to unity when the wavefront is frozen.
In addition, for conceivable wave-control applications based on frozen transmission, such as wavefront filtering and stabilization, the average fraction of energy in the dominant output mode is important: $E=\langle| \mathbf{y}_{\mathcal{R}}^\dagger  \mathbf{u}_1|^2/\lVert \mathbf{y}_{\mathcal{R}} \rVert_2^2\rangle$ with $0\leq E\leq 1$. 

To summarize, the occurrence of frozen wavefronts is associated with a special rank-one structure of the TM that is only found in special types of complex media. 
We emphasize that these frozen wavefronts generally occur \textit{only in transmission}. Indeed, $\mathbf{T}$ being dominated by one singular value does \textit{not} generically imply that $\mathbf{S}$ is also dominated by one singular value. Usually, the reflected wavefront displays the typical sensitivity to the input wavefront even when the transmitted wavefront is frozen.

\subsection{\textit{Differential} scattering in reconfigurable complex media}
\label{subsec_differential_principle}

We now turn our attention to the new phenomenon on which we report in this paper. It applies to scattering in general, including transmission in particular, but also reflection. 
We consider the case of a localized perturbation of a complex medium. The most prominent embodiment thereof is the smart radio environment that is of great contemporary research interest in wireless communications. To physics-consistently capture recurrent scattering, we use a well-established and experimentally verified multiport-network model~\cite{del2025ambiguity}.
An analogous derivation in the coupled-dipole formalism for optical multiple scattering (as used, for instance, in Ref.~\cite{leseur2014probing}) is included in the Supporting Information. 
In certain optical systems like MPLCs, forward scattering is dominant, which simplifies the modeling and allows one to relax the requirement for the electrically small size of the perturbation, as explained in the Supporting Information.

\subsubsection{Multiport-network model} We now briefly review the multiport-network model. Consider a smart radio environment with $N_\mathrm{T}$ transmitting antennas and $N_\mathrm{R}$ receiving antennas. The wireless signal propagation is parametrized by a programmable metasurface comprising $N_\mathrm{S}$ meta-atoms, each endowed with an individually tunable element~\cite{del2025ambiguity}. The model's starting point is a partition of the entire system into three entities: (i) a set of $N_\mathrm{A}=N_\mathrm{T}+N_\mathrm{R}$ antenna ports via which signals are injected and received; (ii) a set of $N_\mathrm{S}$ tunable elements; and (iii) all static scattering objects. The antenna ports and the tunable elements are assumed to be lumped, meaning they are much smaller than the operating wavelength and hence effectively point-like. It is further assumed that all antenna ports are matched, and that the entire system and each of its components is passive and linear. It follows that entity (iii) is characterized by an $N$-port scattering matrix $\check{\mathbf{S}}\in\mathbb{C}^{N \times N}$, where $N=N_\mathrm{A}+N_\mathrm{S}$. Moreover, the $i$th tunable lumped element is modeled as a ``virtual'' lumped port terminated by a load with reflection coefficient $\rho_i\in\mathbb{C}$; it follows that entity (ii) is characterized by an $N_\mathrm{S}$-port diagonal scattering matrix $\mathbf{\Phi} = \mathrm{diag}([\rho_1, \rho_2,\dots,\rho_{N_\mathrm{S}}])\in\mathbb{C}^{N_\mathrm{S}\times N_\mathrm{S}}$. The overall system is the connection of the aforementioned two systems, and hence multiport-network theory yields its $N_\mathrm{A}$-port scattering matrix: $ \mathbf{S} = \check{\mathbf{S}}_\mathcal{AA} + \check{\mathbf{S}}_\mathcal{AS} \left(\mathbf{\Phi}^{-1} - \check{\mathbf{S}}_\mathcal{SS}  \right)^{-1} \check{\mathbf{S}}_\mathcal{SA}\in\mathbb{C}^{N_\mathrm{A}\times N_\mathrm{A}}$. The set $\mathcal{S}$ contains the port indices associated with the tunable elements, and $\mathcal{A}=\mathcal{T}\cup\mathcal{R}$. The parameters of the multiport-network model are generally not readily known, but recent work in wireless communications has established techniques to estimate them accurately~\cite{sol2024experimentally,del2025experimental,del2025ambiguity}.

\subsubsection{Frozen differential wave scattering}
\label{sec_1B2}
Let us consider the case of a single localized perturbation, i.e., $N_\mathrm{S}=1$. In this case, $\check{\mathbf{S}}_\mathcal{AS}$ and $\check{\mathbf{S}}_\mathcal{SA}$ are vectors and $\check{\mathbf{S}}_\mathcal{SS}$ is a scalar, which we denote by $\mathbf{p}$, $\mathbf{q}$, and $g$, respectively, for notational ease. Under reciprocity, $\mathbf{p}=\mathbf{q}^\top$. 
We change the reflection coefficient of the tunable load from $\rho_1$ to $\hat{\rho}_1$. As a result, the system's scattering matrix changes from $\mathbf{S}$ to $\hat{\mathbf{S}} = \mathbf{S} + \mathbf{p} \left( \frac{\hat{\rho}_1}{1-\hat{\rho}_1 g} - \frac{\rho_1}{1-\rho_1 g} \right) \mathbf{q}$. By inspection, we see that the \textit{differential} scattering matrix $\Delta\mathbf{S} = \hat{\mathbf{S}} - \mathbf{S}$ has rank one.\footnote{The rank-one nature of $\Delta\mathbf{S}$ appears as fundamental mathematical ingredient in various recent works~\cite{prod2023efficient,sol2025optimal,del2025virtual,del2025experimental,del2025ambiguity}, but its physical significance in terms of frozen differential scattering was not recognized so far.} Its singular value decomposition thus only has one significant singular value $\sigma_1$ and collapses to $\Delta{\mathbf{S}} = \sigma_1 \mathbf{s}_1 \mathbf{r}_1^\dagger$ ($\mathbf{s}_1$ is collinear with $\mathbf{p}$, and $\mathbf{r}_1$ is collinear with $\mathbf{q}^\dagger$). We emphasize that we have not made any specific assumption about $\mathbf{S}$, meaning that $\mathbf{S}$ generally has many significant singular values. Consequently, the output wavefronts before ($\mathbf{y}=\mathbf{S}\mathbf{x}$) and after ($\hat{\mathbf{y}}=\hat{\mathbf{S}}\mathbf{x}$)  the change are generally sensitive to the choice of $\mathbf{x}$; in other words, they are \textit{not} frozen. However, the shape of the \textit{change} of the output wavefront, i.e., the \textit{differential} output wavefront $\Delta\mathbf{y}=\hat{\mathbf{y}}-\mathbf{y}$, is frozen, being always collinear with $\mathbf{s}_1$. To see this explicitly, we note that $\Delta\mathbf{y}=(\Delta\mathbf{S})\mathbf{x} = \beta(\mathbf{x})\mathbf{s}_1$, where $\beta(\mathbf{x})=\sigma_1 \mathbf{r}_1^\dagger \mathbf{x}$ is a complex-valued scalar. 

Because the perturbation of $\mathbf{S}$ generally affects all entries of $\mathbf{S}$, $\Delta\mathbf{T}$ is also generally of rank one, and $\Delta\mathbf{y}_\mathcal{R}$ is also frozen. Thus, the differential freezing applies to all types of scattering (transmission and/or reflection), whereas the conventional freezing reviewed previously generally only occurs in transmission.

\section{Experimental observation of frozen differential scattering}
\label{sec_experiment}

\begin{figure*}
\centering
\includegraphics[width=0.8\linewidth]{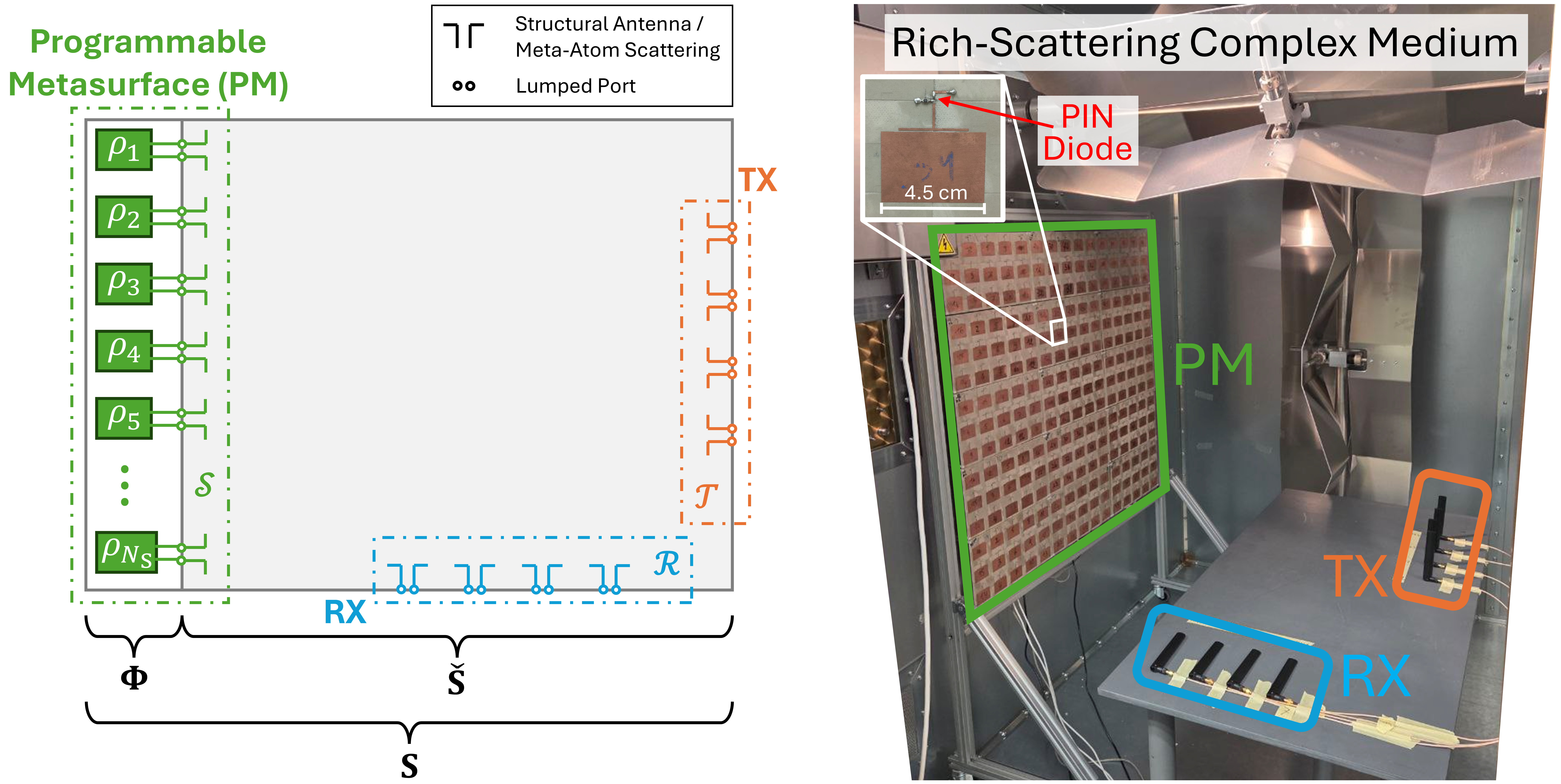}
\caption{Multiport-network schematic (left) and photographic image (right) of the experimental setup involving two times four antennas (TX and RX, cross-polarized) and an array of programmable meta-atoms (programmable metasurface, PM). Each meta-atom consists of a static component (its structural scattering) and a point-like tunable lumped element (PIN diode), as shown in the inset. Changing the configuration of a single meta-atom is a localized perturbation that yields a frozen change of the output wavefront. The remaining meta-atoms are kept in a fixed state (which is optimized for \textit{customized} differential freezing in Sec.~\ref{sec_customized}).}
\label{Fig2}
\end{figure*}

We now report an experimental observation of frozen differential scattering in a smart radio environment. Our experimental setup is shown in Fig.~\ref{Fig2} together with the corresponding multiport-network schematic. $N_\mathrm{A}=8$ antennas allow us to inject/receive waves into/from a reverberation chamber equipped with a 225-element programmable metasurface. The setup emulates a $4\times 4$ MIMO scenario with $N_\mathrm{T}=N_\mathrm{R}=4$. Further experimental details are provided in the Methods. In this section, we only change the state of one meta-atom while keeping all other meta-atoms in a fixed (random) state. We use different realizations of the random state of the 224 other meta-atoms to conveniently access different realizations of our complex system~\cite{ahmed2024over}. Changing the configuration of a single meta-atom constitutes a localized perturbation even though the size of the meta-atom is on the order of half a wavelength. The reason is that only the scattering properties of the PIN diode change, which is a lumped (point-like) element as seen in the inset in Fig.~\ref{Fig2}. It is the PIN diode that we model as a virtual port terminated by a tunable load. Meanwhile, the static components of the meta-atom are part of the background scattering in the complex medium that is characterized by $\check{\mathbf{S}}$. 

Changing the bias voltage of the PIN diode alters the PIN diode's scattering properties across a very broad frequency range, implying a very broadband nature of the localized perturbation. However, the coupling of the PIN diode to the antenna ports is chiefly determined by the static components of the meta-atom surrounding the PIN diode; if the coupling is too weak, measuring the differential scattering in the presence of noise becomes challenging. The mechanism behind the frozen differential scattering phenomenon is hence not frequency-selective but the experimental observability of the freezing is constrained by the meta-atom design.
Our meta-atom relies on a resonant design that covers a range of roughly 100~MHz around the central operating frequency of 2.45~GHz. All experiments reported in this paper are conducted at 2.45~GHz.

\begin{table}[b]
\centering
\caption{Experimentally evaluated freezing metrics $R$, $P$, $C$, and $E$ for $\mathbf{S}$, $\Delta\mathbf{S}$, $\mathbf{T}$, and $\Delta\mathbf{T}$.}
\begin{tabular}{ccccc}
Matrix   & $R$ & $P$ & $C$ & $E$ \\
\midrule
$\mathbf{S}\in\mathbb{C}^{8 \times 8}$        & $6.42\pm0.13$ & $4.05\pm0.11$ & $0.43\pm0.01$ & $0.34\pm0.01$ \\
$\Delta\mathbf{S}\in\mathbb{C}^{8 \times 8}$ & $1.32\pm0.05$ & $1.00\pm0.00$ & $0.99\pm0.00$ & $0.99\pm0.00$ \\
$\mathbf{T}\in\mathbb{C}^{4 \times 4}$ & $3.26\pm0.17$ & $2.11\pm0.17$ & $0.60\pm0.02$ & $0.56\pm0.03$ \\
$\Delta\mathbf{T}\in\mathbb{C}^{4 \times 4}$ & $1.10\pm0.03$ & $1.00\pm0.00$ & $1.00\pm0.00$ & $1.00\pm0.00$ \\
\bottomrule
\label{Table1}
\end{tabular}
\end{table}

To start, we experimentally evaluate the four freezing metrics $R$, $P$, $C$, and $E$. We summarize in Table~\ref{Table1} their mean and standard deviation across 250 random configurations of the 224 other meta-atoms. As expected, neither $\mathbf{S}$ nor $\mathbf{T}$ are dominated by one singular value, so that the output wavefront is not frozen. However, the changes of $\mathbf{S}$ and $\mathbf{T}$ upon reconfiguring the selected meta-atom are dominated by one singular value, as evidenced by all four freezing metrics, suggesting that the differential output wavefront is frozen. This observation confirms that the differential freezing phenomenon on which we report is not limited to transmission. Out of the four freezing metrics, $R$ is seen to be the most stringent one because its value deviates most clearly from unity for both $\Delta\mathbf{S}$ and $\Delta\mathbf{T}$. Of course, in our experimental measurements, the rank of $\Delta\mathbf{S}$ and $\Delta\mathbf{T}$ is only approximately unity because the PIN diode has a very small but finite extent. The lumped-element approximation is very good but not perfect in the experimental reality. Consequently, the non-leading singular values are very small but not exactly zero.

\begin{figure*}
\centering
\includegraphics[width=\linewidth]{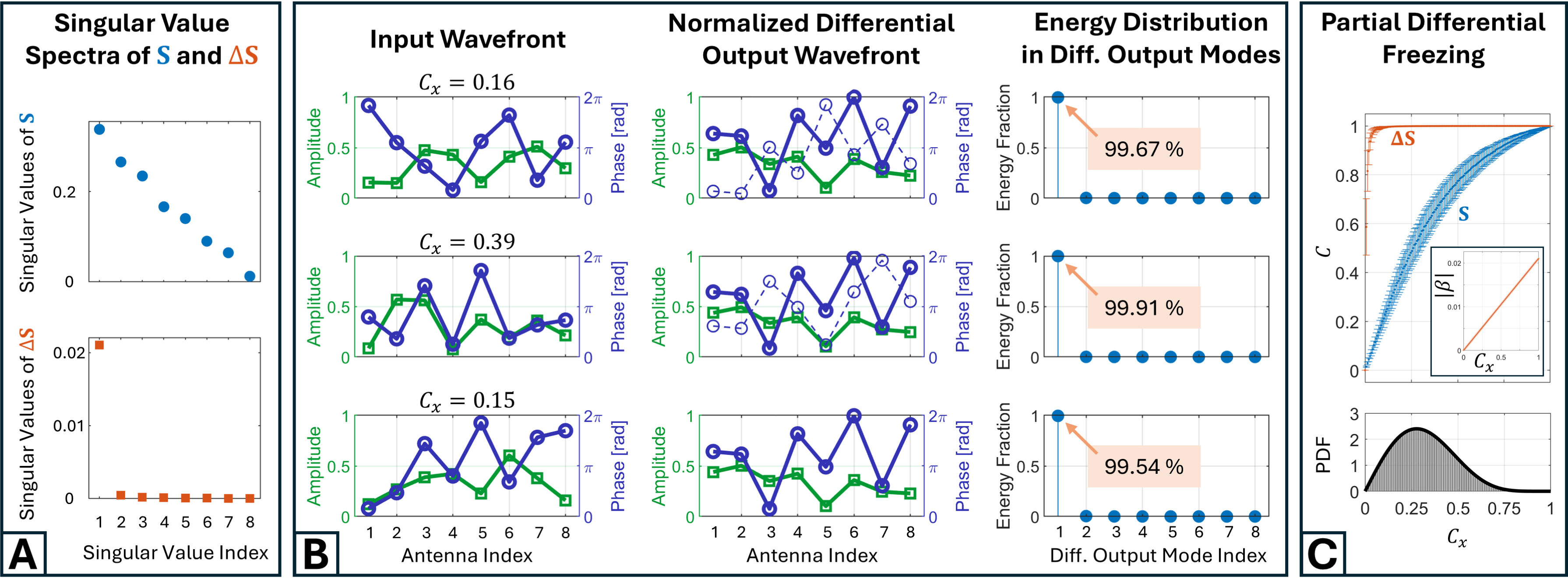}
\caption{Experimental observation of frozen differential scattering; an analogous figure for differential transmission is included as Fig.~\ref{FigS1}.
(A) Measured singular value spectra of $\mathbf{S}$ (top) and $\Delta\mathbf{S}$ (bottom).
(B) For three distinct input wavefronts (one per row), we display three items. 
Left column: $\mathbf{x}$ in terms of amplitude (green, left axis) and phase (blue, right axis); $C_x = | \mathbf{x}^\dagger \mathbf{{r}}_1 | / \lVert \mathbf{x} \rVert_2 \lVert \mathbf{{r}}_1 \rVert_2  $ indicates the overlap of $\mathbf{x}$ with $\mathbf{r}_1$.
Middle column: $\Delta\mathbf{y}/\|\Delta\mathbf{y}\|_2$ in terms of normalized amplitude (green, left axis), aligned phase (blue, right axis), and raw phase (blue dashed, right axis). The alignment consists in applying a global phase rotation $\mathrm{exp}(-\jmath\ \mathrm{arg}(\mathbf{s}_1^\dagger \Delta\mathbf{y}))$ to all entries of $\Delta\mathbf{y}$.
Right column: Energy distribution of the normalized $\Delta\mathbf{y}$ over the output modes (left singular vectors) of $\Delta\mathbf{S}$.
(C) Freezing metric $C$ as a function of $C_x$ for $\Delta\mathbf{S}$ (red), and analogous results for $\mathbf{S}$ (blue). The inset shows the scaling of $|\beta|$ with $C_x$. The probability density function (PDF) of $C_x$ for random wavefronts is also shown.
}
\label{Fig3}
\end{figure*}

Because $\Delta\mathbf S$ is experimentally very close—but not exactly—rank one, small deviations from perfect freezing of the differential output wavefront can occur. Writing
$\Delta\mathbf y = \beta \mathbf s_1 + \sum_{k>1}\sigma_k(\mathbf r_k^\dagger\mathbf x)\,\mathbf s_k$, we see that the differential output wavefront is the frozen shape $\mathbf s_1$ scaled by $\beta$  plus leakage from subdominant singular directions. As seen in the inset of Fig.~\ref{Fig3}C, $|\beta|$ is proportional to the collinearity of $\mathbf{x}$ and $\mathbf{r}_1$, which we refer to as 
$C_x = |\mathbf x^\dagger \mathbf r_1|$ (with $\|\mathbf x\|_2=1$). 
For random input wavefronts, $C_x$ is rarely extremely small (Fig.~\ref{Fig3}C, bottom), so the subdominant terms are strongly suppressed by $\sigma_{k>1}\!\ll\!\sigma_1$. Consequently, the freezing metric $C$ stays close to unity as long as $C_x$ does not approach zero, as seen in Fig.~\ref{Fig3}C. Only in the pathological limit $C_x\!\to\!0$ would the differential output wavefront cease to align with $\mathbf s_1$, but then $|\beta|\!\to\!0$ and the differential signal vanishes, making this unlikely event practically irrelevant. In contrast, for total scattering the freezing metric stays close to unity only if input wavefronts that are almost perfectly collinear with the first right singular vector of $\mathbf{S}$ are chosen (blue curve in Fig.~\ref{Fig3}C). In other words, total scattering displays the usual sensitivity to the input wavefront in our complex medium.

Next, for one arbitrary realization of the complex medium, we explicitly observe the frozen differential scattering in Fig.~\ref{Fig3}B. The singular value spectra of $\mathbf{S}$ and $\Delta\mathbf{S}$ in Fig.~\ref{Fig3}A visualize the dominance of one singular value in differential scattering but not in total scattering, in line with our observations in Table~\ref{Table1}. In Fig.~\ref{Fig3}B, we consider three random normalized input wavefronts. They are displayed in the left column of Fig.~\ref{Fig3}B; their overlap $C_x$ with the dominant right singular vector of $\Delta\mathbf{S}$ is indicated and seen to be small. In the middle column of Fig.~\ref{Fig3}B, we display the corresponding normalized differential output wavefronts associated with reconfiguring the selected meta-atom. Upon visual inspection, the amplitudes and aligned phases are extremely similar, providing explicit evidence for the frozen differential scattering. The phase alignment refers to adjusting a global phase factor to facilitate visual comparison, as detailed in the caption of Fig.~\ref{Fig3}. In the right column of Fig.~\ref{Fig3}B, we show the corresponding energy distributions in the basis of the differential output modes. The latter are the left singular vectors of $\Delta\mathbf{S}$~\cite{miller2019waves}. In all three cases, the dominant mode carries more than 99\ \% of the energy.

A figure analogous to Fig.~\ref{Fig3} but for differential transmission as opposed to differential scattering is provided in the Supporting Information.

\section{Partial coherence and thermal emission}

Existing works on frozen wavefronts, as well as our present work thus far, assume that, \textit{first}, the input wavefront $\mathbf{x}$ is perfectly coherent and, \textit{second}, the complex medium does not generate thermal emission. In this section, we generalize the theories of frozen transmission and frozen differential scattering to cases with partially coherent impinging wavefronts and thermal emission by the complex medium.

The coherence of the impinging wavefront $\mathbf{x}$ is characterized by its coherence matrix $\mathbf{\Gamma}_\mathrm{in} = \langle \mathbf{x} \mathbf{x}^\dagger \rangle \in\mathbb{C}^{{N_\mathrm{A}}\times {N_\mathrm{A}}}$, where the average is over realizations of the statistical processes. If the incident wavefront is perfectly coherent, $\mathbf{\Gamma}_\mathrm{in}$ is of rank one. If thermal emission by the complex medium is negligible, the coherence matrix $\mathbf{\Gamma}_\mathrm{out} = \langle \mathbf{y} \mathbf{y}^\dagger \rangle \in\mathbb{C}^{{N_\mathrm{A}}\times {N_\mathrm{A}}}$ of the outgoing wavefront is related to $\mathbf{\Gamma}_\mathrm{in}$ and $\mathbf{S}$ as follows: $\mathbf{\Gamma}_\mathrm{out} = \mathbf{S} \mathbf{\Gamma}_\mathrm{in} \mathbf{S}^\dagger$~\cite{wolf1982new,wolf1986new,withington2002modal,zhang2019scattering,roques2024measuring,guo2024unitary,guo2024unitary2,harling2025optical}. Taking thermal emission into account yields $\mathbf{\Gamma}_\mathrm{tot} = \mathbf{\Gamma}_\mathrm{out}+\mathbf{\Gamma}_\mathrm{th}$, where $\mathbf{\Gamma}_\mathrm{th}\in\mathbb{C}^{{N_\mathrm{A}}\times {N_\mathrm{A}}}$ represents thermal emission from the complex medium and is itself related to $\mathbf{S}$ as follows: $\mathbf{\Gamma}_\mathrm{th}= \theta\big(\mathbf{I}_{N_\mathrm{A}}-\mathbf S \mathbf S^\dagger\big) $, with $\theta$ being a scaling factor that depends on temperature and frequency~\cite{twiss1955nyquist,bosma1967theory,haus2012electromagnetic,wedge2002noise,miller2017universal}. 
This expression assumes that the complex medium is passive and in isothermal equilibrium, and that its ports are matched. Note that in the limit in which the complex system is lossless and hence characterized by a unitary $\mathbf{S}$, the emission of thermal noise vanishes: $\mathbf{\Gamma}_\mathrm{th}=\mathbf{0}$. For transmission-only settings, it follows that $\mathbf{\Gamma}_{\mathrm{tot},\mathcal{R}\mathcal{R}}
=\mathbf T\,\mathbf{\Gamma}_{\mathrm{in},\mathcal{T}\mathcal{T}}\,\mathbf T^\dagger
+\mathbf{\Gamma}_{\mathrm{th},\mathcal{R}\mathcal{R}}$. 

Altogether, the interaction of the wavefront with the complex medium generally alters its coherence matrix. 
It is hence intriguing to ask whether the frozen wavefront phenomenon, both in its original embodiment in transmission and our new embodiment in differential scattering, imposes any particular structure on $\mathbf{\Gamma}_\mathrm{tot}$ and/or its components, or their differential counterparts.

\subsection{Frozen transmission}
If thermal emission is negligible, $\mathbf{\Gamma}_{\mathrm{out},\mathcal{R}\mathcal{R}}
=\mathbf{T} \mathbf{\Gamma}_{\mathrm{in},\mathcal{T}\mathcal{T}} \mathbf{T}^\dagger
\approx \tau_1^2 \mathbf{u}_1\big(\mathbf{v}_1^\dagger \mathbf{\Gamma}_{\mathrm{in},\mathcal{T}\mathcal{T}} \mathbf{v}_1\big)\mathbf{u}_1^\dagger = \nu \mathbf{u}_1 \mathbf{u}_1^\dagger$, where $\nu = \tau_1^2 \mathbf{v}_1^\dagger \mathbf{\Gamma}_{\mathrm{in},\mathcal{T}\mathcal{T}} \mathbf{v}_1$ is a real-valued, non-negative scalar. It is apparent that $\mathbf{\Gamma}_{\mathrm{out},\mathcal{R}\mathcal{R}}$ is a rank-one matrix regardless of the rank of $\mathbf{\Gamma}_{\mathrm{in},\mathcal{T}\mathcal{T}}$. Thus, \textit{frozen wavefronts are always perfectly coherent}, irrespective of the degree of coherence of the input wavefront. The coherence of the input wavefront only affects a global scaling factor. A complex medium featuring frozen transmission is hence a (possibly highly non-trivial) coherence purifier.

Let us now turn our attention to cases with thermal emission from the complex medium. The thermal noise power emitted into a spatial output mode $\mathbf{w}$ (where $\mathbf{w}\in\mathbb{C}^{N_\mathrm{R}}$ is a unit vector) is $\mathbf{w}^\dagger \mathbf{\Gamma}_{\mathrm{th},\mathcal{RR}}\mathbf{w} = \theta\!\left( 1
-\big\|\mathbf S_{\mathcal{RR}}^\dagger \mathbf w\big\|_2^2
-\big\|\mathbf T^\dagger \mathbf w\big\|_2^2 \right)$. For a complex medium with frozen transmission, the term $\big\|\mathbf T^\dagger \mathbf w\big\|_2^2\approx\tau_1^2|\mathbf u^\dagger_1 \mathbf w|^2$ implies that the more a spatial mode $\mathbf{w}$ overlaps with $\mathbf{u}_1$, the more the thermal noise emission into $\mathbf{w}$ is suppressed. This suppression is thus maximized if we choose $\mathbf{w}$ to be the spatial mode $\mathbf{u}_1$ in which the transmission is frozen. Consequently, \textit{thermal noise is partially suppressed in frozen transmission}, facilitating the separation of signal and noise.

Two special cases are particularly insightful. 
\textit{First}, assuming $\mathbf S_{\mathcal{RR}}$ is a random matrix independent of $\mathbf{u}_1$, the expected value of the noise power in the frozen transmission mode is $\eta_\mathrm{f} = \eta-\theta\tau_1^2$, where $\eta = \theta(1-\big\|\mathbf S_{\mathcal{RR}} \big\|_\mathrm{F}^2/N_\mathrm{R})$ is the expected value of the noise power in any spatial mode orthogonal to $\mathbf{u}_1$ (by isotropy, they are all equal on average). 
\textit{Second}, assuming $\mathbf S_{\mathcal{RR}}=\mathbf{0}$, it follows that $\eta=\theta$ and $\eta_\mathrm{f}=\theta(1-\tau_1^2)$, implying that thermal emission into the frozen transmission's spatial mode can be totally suppressed as $\tau_1\rightarrow 1$. 
We have three remarks for completeness: (i) $\mathbf S_{\mathcal{RR}}=\mathbf{0}$ is routinely assumed in widespread models of 
MPLCs (see Supporting Information). (ii) $\mathbf S_{\mathcal{RR}}=\mathbf{0}$ implies that the complex medium has numerous reflectionless scattering modes, which can only be the case at discrete frequencies but not broadband~\cite{sweeney2020theory}. (iii) If the complex medium additionally has some coherent perfect absorption modes, then the associated blackbody thermal radiation is orthogonal to the frozen transmission and thus physically separable~\cite{ona2024orthogonal}.

To summarize, frozen transmission purifies coherence, and the emission of thermal noise into the spatial mode of the frozen transmission is partially suppressed. These insights bring a fresh perspective to the physics of frozen transmission, relevant to coherence and thermal noise management.

\subsection{Frozen differential scattering}
\label{sec_3B}

If thermal emission is negligible, the differential output wavefront $\Delta\mathbf{y}=(\Delta \mathbf{S})\mathbf{x}$ is always perfectly coherent: $\mathbf{\Gamma}_\mathrm{out,\Delta}=\langle \Delta\mathbf{y} (\Delta\mathbf{y})^\dagger \rangle = (\Delta\mathbf{S}) \mathbf{\Gamma}_\mathrm{in} (\Delta\mathbf{S})^\dagger = \gamma \mathbf{s}_1 \mathbf{s}_1^\dagger$, where $\gamma = \sigma_1^2 \mathbf{r}_1^\dagger \mathbf{\Gamma}_\mathrm{in} \mathbf{r}_1$ is a real-valued, non-negative scalar. Applications based on such differential output fields thus do \textit{not} require any coherence of the input wavefront, making them compatible with the use of partially or purely incoherent sources.
Moreover, the change of the total output wavefront's coherence matrix is at most of rank two. Indeed, the row and column spaces of $\Delta\mathbf{\Gamma}_\mathrm{out}$ lie inside a 2D-span defined by $\mathbf{s}_1$ and $\mathbf{a}=\sigma_1\mathbf{S}\mathbf{\Gamma}_\mathrm{in}\mathbf{r}_1\in\mathbb{C}^{N_\mathrm{A}}$: $\Delta\mathbf{\Gamma}_\mathrm{out} = \mathbf{s}_1\mathbf{a}^\dagger + \mathbf{a}\mathbf{s}_1^\dagger + \mu\mathbf{s}_1\mathbf{s}_1^\dagger$, where $\mu = \sigma_1^2\mathbf{r}_1^\dagger\mathbf{\Gamma}_\mathrm{in}\mathbf{r}_1$ is a real-valued, non-negative scalar. 
Analogously, $\Delta\mathbf{y}_\mathcal{R}$ is perfectly coherent and $\Delta\mathbf{\Gamma}_{\mathrm{out},\mathcal{RR}}$ is at most of rank two. 

Let us now turn our attention to thermal noise emission from the complex medium. We assume that the thermal noise realizations before and after the rank-one perturbation of the complex medium are independent: $\langle {\mathbf{n}} \hat{\mathbf{n}}^\dagger  \rangle=\mathbf{0}$. We begin by examining how the perturbation affects the thermal noise coherence; then, we derive an expression for the signal-to-thermal-noise ratio (STNR) in differential scattering measurements with awareness of the frozen shape of  $\Delta\mathbf{y}$.

The change of the coherence matrix of the total thermal emission is  $\Delta\mathbf{\Gamma}_\mathrm{th} = -\theta\big(\mathbf s_1\mathbf b^\dagger + \mathbf b\,\mathbf s_1^\dagger + \sigma_1^2\,\mathbf s_1\mathbf s_1^\dagger\big) $, where $\mathbf{b}=\sigma_1\,\mathbf S\,\mathbf r_1\in\mathbb C^{{N_\mathrm{A}}}$. We see by inspection that the row and column spaces of $\Delta\mathbf{\Gamma}_\mathrm{th}$ lie inside a 2D-span defined by $\mathbf{s}_1$ and $\mathbf{b}$, implying that the coherence matrix of the thermal emission changes at most by rank two. This span shares one direction (namely $\mathbf{s}_1$) with the span of $\Delta\mathbf{\Gamma}_\mathrm{out}$ derived in the previous paragraph in the absence of thermal noise. Thus, in the presence of thermal noise, $\Delta\mathbf{\Gamma}_\mathrm{tot}$ is generally at most of rank 3 with a 3D-span defined by $\mathbf{s}_1$, $\mathbf{a}$, and $\mathbf{b}$.
In a transmission setup, analogous results can be derived for $\Delta\mathbf{\Gamma}_{\mathrm{th},\mathcal{RR}}$ and $\Delta\mathbf{\Gamma}_{\mathrm{tot},\mathcal{RR}}$.

Now, we derive the STNR for measurements of the signals in the change of the output wavefront. Such measurements boil down to estimating a scalar signal $z$ that is the projection of $\Delta\mathbf{y}$ along $\mathbf{s}_1$. We have $z=\sigma_1\mathbf{r}_1^\dagger \mathbf{x} + \mathbf{s}_1^\dagger  (\hat{\mathbf{n}}-\mathbf{n}) $, where the first term is the desired signal and the second term is the thermal noise.
The signal power is thus $\sigma_1^2\mathbf{r}_1^\dagger\mathbf{\Gamma}_\mathrm{in}\mathbf{r}_1$; the thermal noise power is $\mathbf{s}_1^\dagger \langle (\hat{\mathbf{n}}-\mathbf{n}) (\hat{\mathbf{n}}-\mathbf{n})^\dagger \rangle = \theta \left(2-\big\|\mathbf S^\dagger \mathbf{s}_1\big\|_2^2 -\big\|\hat{\mathbf{S}}^\dagger {\mathbf{s}}_1\big\|_2^2 \right)$. The more the dominant left singular vectors of $\mathbf{S}$ and $\hat{\mathbf{S}}$ are collinear with $\mathbf{s}_1$, the less thermal noise affects our frozen differential measurement. Moreover, the signal power can be maximized by choosing $\mathbf{\Gamma}_\mathrm{in}$ such that its dominant eigenvector is maximally aligned with $\mathbf{r}_1$. In a transmission setup, analogous  expressions can be derived.

To summarize, frozen differential scattering imposes significant structure on the associated changes in coherence. Specifically, the perturbation's rank-one nature imposes bounds on the possible changes in coherence. Exploiting the latter's low-rank nature can be particularly valuable when the dimensions of $\mathbf{S}$ (or $\mathbf{T}$) are large, as in massive-MIMO wireless systems and optical wavefront shaping with hundreds of inputs and outputs. Several applications can be envisioned, including, low-complexity measurements of the differential coherence matrix inspired by the approach in Ref.~\cite{roques2024measuring} or based on compressed-sensing principles, efficient denoising, or schemes for covert and nearly-passive communications with purposefully perturbed thermal noise.

\section{Customized differential freezing}
\label{sec_customized}

So far, we considered a single localized rank-one perturbation and showed that the associated differential scattering is frozen. While its frozen nature is intriguing, the specific shape in which it freezes is generally seemingly arbitrary in a complex medium. An arbitrarily shaped mode can be inconvenient for measurements or applications; instead,  a one-hot pattern or a spatially uniform pattern may be preferable. Beyond prescribing the pattern of the frozen differential output wavefront, experimental schemes can also benefit from maximizing the STNR in freezing-aware measurements of differential signals. 
Given these motivations, we now seek to \emph{customize} the frozen differential scattering in terms of the frozen differential output shape or the STNR. For concreteness, we consider frozen differential transmission because in many applications transmission measurements are easier and more common than reflection measurements.

Customizing the frozen differential scattering requires additional control over the complex medium. We thus now consider a complex medium featuring an additional set $\mathcal{X}$ of $N_\mathrm{X}$ independently tunable localized perturbations. A contemporary embodiment thereof is a smart radio environment, wherein the utilized programmable metasurface is an array of individually programmable meta-atoms, each endowed with an individually controllable lumped element that acts like a localized rank-one perturber.  Our goal is to optimize the configuration of the programmable meta-atoms in $\mathcal{X}$ to tune the complex medium to a state in which changing the remaining meta-atom's configuration yields a differential output wavefront of the desired shape or with enhanced STNR. Our setup is hence exactly the one shown in Fig.~\ref{Fig2}, but in contrast to Sec.~\ref{sec_experiment} we optimize the configuration of the 224 remaining meta-atoms instead of using a random configuration.

\subsection{Principle}

We proceed with formalizing this goal. The set of virtual port indices is now partitioned: $\mathcal{S}=\{\xi\}\cup\mathcal{X}$, where $\xi$ is the \emph{single} remaining port we will vary to induce the rank-one perturbation, and the ports in $\mathcal X$ will be held fixed at optimized terminations to shape the frozen response. 
Let $\mathbf\Omega\in\mathbb C^{N_2\times N_2}$, with $N_2=N_\mathrm A+1+N_\mathrm X$, be the scattering matrix of the static parts of the overall system indexed by $\bar{\mathcal A}\cup\mathcal X$, where $\bar{\mathcal A}=\mathcal A\cup\{\xi\}$. 
The $i$th virtual port in $\mathcal X$ is terminated by an individual load with reflection coefficient $\bar\rho_i$, and we define $\bar{\mathbf\Phi}=\mathrm{diag}([\bar\rho_1,\ldots,\bar\rho_{N_\mathrm X}])\in\mathbb C^{N_\mathrm X\times N_\mathrm X}$. Thus, by multiport-network theory, the resulting $(N_\mathrm A+1)$-port system whose ports are those in $\bar{\mathcal A}$ is characterized by the scattering matrix $\check{\mathbf S}(\bar{\mathbf\Phi})
=\mathbf\Omega_{\bar{\mathcal A}\bar{\mathcal A}}
+\mathbf\Omega_{\bar{\mathcal A}\mathcal X}\big(\bar{\mathbf\Phi}^{-1}-\mathbf\Omega_{\mathcal X\mathcal X}\big)^{-1}\mathbf\Omega_{\mathcal X\bar{\mathcal A}}$.
$\check{\mathbf S}(\bar{\mathbf\Phi})$ plays exactly the same role as before, only now it depends on the chosen terminations of $\mathcal X$ and can thus be customized.

Analogously to before, but now specialized to transmission, we have  $\Delta{\mathbf{T}}(\bar{\mathbf\Phi}) = \breve{\mathbf{p}}(\bar{\mathbf\Phi}) \left( \frac{\hat{\rho}_1}{1-\hat{\rho}_1 g(\bar{\mathbf\Phi})} - \frac{\rho_1}{1-\rho_1 g(\bar{\mathbf\Phi})} \right) \breve{\mathbf{q}}(\bar{\mathbf\Phi})$, where $\breve{\mathbf{p}}\in\mathbb{C}^{N_\mathrm{R}}$, $\breve{\mathbf{q}}\in\mathbb{C}^{N_\mathrm{T}}$, and we emphasize the new dependence on $\bar{\mathbf{\Phi}}$. For clarity, we use the same variable names as for the scattering analysis in Sec.~\ref{sec_1B2}, but we indicate with a breve symbol on top our specialization to transmission.
Applying an SVD yields $\Delta{\mathbf{T}}(\bar{\mathbf\Phi}) = \breve{\sigma}_1(\bar{\mathbf\Phi}) \breve{\mathbf{s}}_1(\bar{\mathbf\Phi}) \breve{\mathbf{r}}_1^\dagger(\bar{\mathbf\Phi})$, where $\breve{\mathbf s}_1(\bar{\mathbf\Phi})=\breve{\mathbf p}(\bar{\mathbf\Phi})/\|\breve{\mathbf p}(\bar{\mathbf\Phi})\|_2$ is the frozen shape of the differential output wavefront, and $\breve{\mathbf r}_1(\bar{\mathbf\Phi})=\breve{\mathbf q}(\bar{\mathbf\Phi})^\dagger/\|\breve{\mathbf q}(\bar{\mathbf\Phi})\|_2$. Now, we can express our optimization objective (e.g., regarding the shape of $\breve{\mathbf{s}}_1$) in terms of the configuration of the additional programmable meta-atoms encoded in $\bar{\mathbf{\Phi}}$. 

Although we know in principle how to estimate the multiport-network model parameters for our experimental setup~\cite{sol2024experimentally,del2025experimental,del2025ambiguity}, we opt for a model-agnostic optimization of $\bar{\mathbf{\Phi}}$ here because the focus in this section is on the optimization outcome rather than the optimization method; a model-based optimization would thus be unnecessarily complicated due to the required parameter estimation. Details of our cost definition and model-agnostic optimization algorithm under the binary constraints of our programmable-metasurface prototype are provided in the Methods.

\subsection{Experimental Demonstration}

\begin{figure}
\centering
\includegraphics[width=0.5\linewidth]{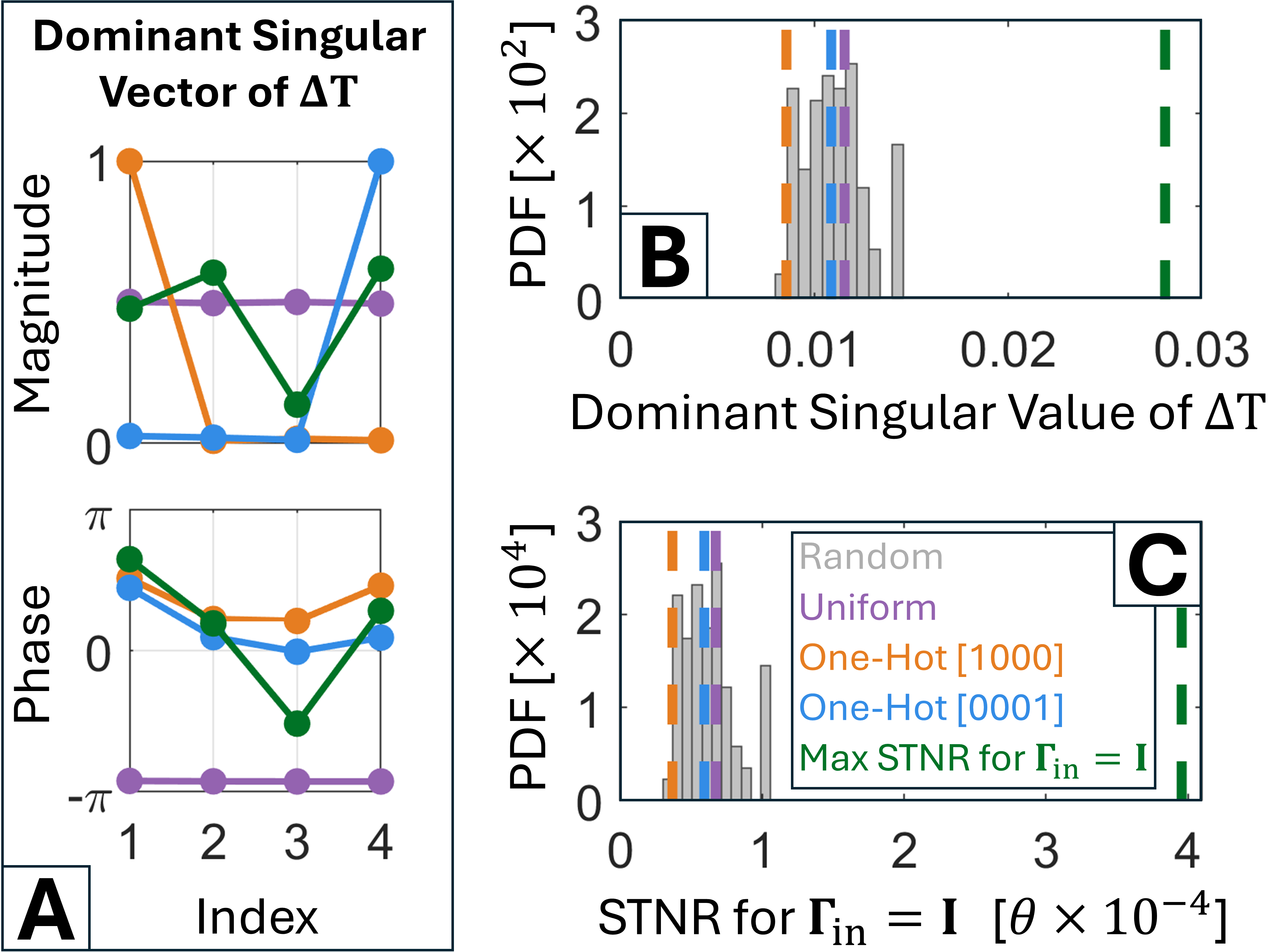}
\caption{Experimental demonstration of customized differential freezing in transmission. 
(A) Measured dominant singular vector of $\Delta\mathbf{T}$ for four optimized configurations of the 224 other meta-atoms; the colors are defined in the legend in (C).
(B) PDF of the dominant singular value of $\Delta\mathbf{T}$ for 250 random configurations of the 224 other meta-atoms; the values corresponding to the four optimized configurations are indicated.
(C) Same as in (B) for the STNR for $\mathbf{\Gamma}_{\mathrm{in},\mathcal{TT}}=\mathbf{I}_4$.
} 
\label{Fig4}
\end{figure}

We begin by presenting in Fig.~\ref{Fig4}A the experimentally measured $ \breve{\mathbf{s}}_1$ after four selected optimizations of $\bar{\mathbf{\Phi}}$ for four selected customizations of differential freezing. For the three examples concerned with the shape of $ \breve{\mathbf{s}}_1$ (uniform: purple; one-hot: orange and blue), we can visually confirm that our optimization outcome closely resembles the desired one. In the case of a uniform $ \breve{\mathbf{s}}_1$ both the magnitude and the phase are almost identical for each entry of  $ \breve{\mathbf{s}}_1$. In the case of a one-hot  $ \breve{\mathbf{s}}_1$, all entries of  $ \breve{\mathbf{s}}_1$ vanish except one (consequently, phases are irrelevant). Incidentally, we note that the one-hot cases are also relevant to covert passive back-scatter communications, where only one intended receive antenna can detect the transmitted information. For the fourth example (green), we sought to maximize the STNR of the differential transmission for $\mathbf{\Gamma}_{\mathrm{in},\mathcal{TT}}=\mathbf{I}_4$. In this case, visual inspection of $\breve{\mathbf{s}}_1$ after optimization does not reveal any particular pattern.

Next, we examine how our optimization of the shape or STNR impacts the dominant singular value $\breve{\sigma}_1$ of $\Delta\mathbf{T}$. In Fig.~\ref{Fig4}B, we superpose the values corresponding to the four optimizations over a PDF of the distribution of $\breve{\sigma}_1$ across 250 random configurations. We observe that the three shape optimizations do not result in a significant increase or decrease of $\breve{\sigma}_1$, meaning the differential signal strength after the shape optimization is comparable to before the shape optimization. Meanwhile, we see that in the case of the STNR maximization, the value of $\breve{\sigma}_1$ is 2.5 times larger than the mean over random configurations. This makes sense because a stronger differential signal boosts the STNR. Indeed, we derived in Sec.~\ref{sec_3B} that the STNR is proportional to $\breve{\sigma}_1^2$.

However, optimizing the STNR is generally not equivalent to maximizing $\breve{\sigma}_1^2$ because the STNR also depends on  $\breve{\mathbf{r}}_1^\dagger\mathbf{\Gamma}_{\mathrm{in},\mathcal{TT}}\breve{\mathbf{r}}_1$, $\big\|\mathbf T^\dagger \breve{\mathbf{s}}_1\big\|_2^2$, and $\big\|\hat{\mathbf{T}}^\dagger \breve{\mathbf{s}}_1\big\|_2^2$. Importantly, the STNR depends hence on $\mathbf{\Gamma}_{\mathrm{in},\mathcal{TT}}$ so STNR maximization is specific to $\mathbf{\Gamma}_{\mathrm{in},\mathcal{TT}}$. We chose $\mathbf{\Gamma}_{\mathrm{in},\mathcal{TT}}=\mathbf{I}_4$ which corresponds to a perfectly incoherent, spatially isotropic input wavefront.
In Fig.~\ref{Fig4}C, we superpose the values corresponding to the four optimizations over a PDF of the distribution of the STNR for $\mathbf{\Gamma}_{\mathrm{in},\mathcal{TT}}=\mathbf{I}_4$ across 250 random configurations. We see that our  configuration optimized to maximize the STNR achieved a 6.3-fold improvement in STNR compared to the mean STNR over random configurations.

Altogether, we have shown in this section that we can customize the spatial pattern of the differential output signal without loss of STNR and that we can maximize the STNR. Combinations of these optimization goals and alternative optimization objectives are left for future work.

\section{Conclusion}

To summarize, we have unveiled the phenomenon of \textit{frozen differential scattering} in reconfigurable complex media. Specifically, a sufficiently localized perturbation of a complex medium results in a change of the output wavefront whose spatial pattern is frozen no matter the input wavefront. We experimentally observed this phenomenon and examined its influence on coherence and thermal noise emission. Then, we customized the frozen differential scattering to impose a specific pattern or to maximize the STNR. 
Our results deepen the understanding of wave scattering in reconfigurable complex media, including situations with partial coherence and thermal emission, and can be exploited for optimal read-out strategies in differential wave-scattering measurements. We also envision applications in filtering and stabilizing differential wavefronts, and expect our insights to benefit imaging, sensing, and communication in reconfigurable complex media.

\section*{Methods}

\subsection*{Notation}
$\mathbf{A} = \mathrm{diag}(\mathbf{a})$ denotes the diagonal matrix $\mathbf{A}$ whose diagonal entries are $\mathbf{a}$. 
$\mathbf{A}_\mathcal{BC}$ denotes the block of the matrix $\mathbf{A}$ whose row [column] indices are in the set $\mathcal{B}$ [$\mathcal{C}$]. 
$^\top$ and $^\dagger$ denote the transpose and transpose-conjugate operations, respectively.
$\lVert\cdot\rVert_2$ denotes the Euclidean (2-) norm for vectors and the induced operator 2-norm for matrices.
$\lVert\cdot\rVert_\mathrm{F}$ denotes the Frobenius norm for matrices.
$\mathbf{I}_a$ denotes the $a\times a$ identity matrix.
$\mathbf{0}$ denotes a matrix (with dimensions determined by context) whose entries are all zeros.
$\jmath$ denotes the imaginary unit.

\subsection*{Experimental setup and procedure}

We conduct our experiments inside a $1.75\ \mathrm{m}\times1.5\ \mathrm{m}\times2\ \mathrm{m}$ reverberation chamber with irregular geometry. The mode stirrers seen in Fig.~\ref{Fig2} are not used and remain static throughout all experiments. 
The reverberation chamber is a complex medium that is routinely used to emulate rich-scattering wireless communication links, as well as for electromagnetic compatibility tests.
Our transmit (TX) and receive (RX) arrays each comprise four identical, commercial WiFi antennas (ANT-W63WS2-SMA) that are regularly spaced in a line by half the operating wavelength. We thus have $N_\mathrm{T}=N_\mathrm{R}=4$, and $N_\mathrm{A}=8$. The TX and RX arrays are oriented perpendicular to each other to minimize the influence of line-of-sight paths that do not interact with the metasurface. Using an eight-port vector network analyzer (VNA), specifically two cascaded Keysight P5024B 4-port
VNAs, we directly measure the $8\times 8$ scattering matrix $\mathbf{S}$ of our system.

The programmable meta-atom is designed for operation at 2.45~GHz and it is based on two tightly coupled resonators: a patch resonator and a parasitic resonator. A PIN diode is integrated into the parasitic resonator. Depending on the applied bias voltage, the PIN diode is conducting or isolating, which affects the effective length and resonance frequency of the parasitic resonator, and thus the overall response of the meta-atom. The meta-atom size, as well as the spacing between adjacent meta-atoms, is roughly half the operating wavelength. The programmable meta-atoms act on a single field polarization whose reflection coefficient phases under normal incidence differ by roughly $\pi$ between the two possible states.
Further details on geometric, material, and electromagnetic aspects of the meta-atom can be found in Ref.~\cite{ahmed2024over} and references therein. 
Our present work does not require detailed knowledge of the meta-atom design.
For the purposes of the present paper, the only pivotal technical detail is that the PIN diode is electrically small. All static parts of the meta-atoms are lumped into $\check{\mathbf{S}}$ while the tunable loads of the meta-atoms are captured by $\mathbf{\Phi}$. The size of the meta-atom being half the operating wavelength, and thus not point-like, is hence not a problem for our assumption of a localized perturbation.

Using Arduino microcontrollers, we can independently adjust the bias voltage of each meta-atom. In total, our metasurface comprises 225 meta-atoms.

\subsection*{Optimization for customized freezing}
We first define a cost $\mathcal{C}$ to be minimized by tuning the configuration of the programmable meta-atoms in $\mathcal{X}$. 

To impose a specific shape of the frozen differential output wavefront, we define $\mathcal{C}$ as the negative of the collinearity of the current $\breve{\mathbf{s}}_1$ with a target vector $\mathbf{t}$: $\mathcal{C} =  -| \mathbf{t}^\dagger \breve{\mathbf{s}}_1 | / \lVert \mathbf{t} \rVert_2 \lVert \breve{\mathbf{s}}_1 \rVert_2 $. For a one-hot shape we use, for example, $\mathbf{t}=[1,0,0,0]$, and for a uniform shape we use $\mathbf{t}=[1,1,1,1]/2$.

To maximize the STNR of a freezing-aware measurement of the transmitted differential signal, we define  $\mathcal{C}$ as the negative of the STNR. In differential transmission, the STNR is the ratio of the signal power $\breve{\sigma}_1^2\breve{\mathbf{r}}_1^\dagger\mathbf{\Gamma}_{\mathrm{in},\mathcal{TT}}\breve{\mathbf{r}}_1$ and the thermal noise power $\theta \left(2-\big\|\mathbf T^\dagger \breve{\mathbf{s}}_1\big\|_2^2 -\big\|\hat{\mathbf{T}}^\dagger \breve{\mathbf{s}}_1\big\|_2^2 \right)$.
Importantly, the signal power, and hence the STNR, depends on $\mathbf{\Gamma}_{\mathrm{in},\mathcal{TT}}$. For the optimization reported in Fig.~\ref{Fig4}, we have chosen $\mathbf{\Gamma}_{\mathrm{in},\mathcal{TT}}=\mathbf{I}_4$ which corresponds to a perfectly incoherent, spatially isotropic input wavefront.

For a given $\mathcal{C}$, we perform a discrete optimization of the configuration of the programmable meta-atoms in $\mathcal{X}$ using coordinate descent, which was shown to be the most efficient approach under these conditions~\cite{hammami2025statistical}. For simplicity, we implement the coordinate descent in a model-agnostic manner by conducting measurements for each iteration. Specifically, we begin by identifying the lowest-cost configuration among 250 random configurations. Using this configuration as initialization, we consider each programmable meta-atom in $\mathcal{X}$ in turn and check whether flipping its state lowers the cost. We loop multiple times over all programmable meta-atoms until convergence. This is necessary due to the non-linear nature of the dependence of the cost on the configuration, originating from multiple scattering and usually also the definition of the cost based on $\check{\mathbf{S}}$. We stop after an entire loop over all programmable meta-atoms without updating.

\section*{Acknowledgements}

This work was supported in part by the ANR France 2030 program (project ANR-22-PEFT-0005), the ANR PRCI program (project ANR-22-CE93-0010), the Rennes M\'etropole AES program (project ``SRI''), the European Union's European Regional Development Fund, and the French Region of Brittany and Rennes M\'etropole through the contrats de plan \'Etat-R\'egion program (projects ``SOPHIE/STIC \& Ondes'' and ``CyMoCoD'').
The author acknowledges I.~Ahmed, F. Boutet, and C. Guitton, who, under the author's supervision, previously built the programmable metasurface prototype for the work presented in Ref.~\cite{ahmed2024over}.

\section*{Supporting Information}

\subsection*{Frozen differential scattering in the coupled-dipole model}

The coupled-dipole formalism~\cite{lax1952multiple} can be formulated for problems involving input and output channels that are localized in space or in momentum space. The former is operationally equivalent to the multiport-network model in the main text~\cite{sol2025optimal,sol2024experimentally}. Here, we detail the coupled-dipole formalism for a 2D setup with channels localized in momentum space, similar to the one considered in Ref.~\cite{leseur2014probing}. However, as in Ref.~\cite{leseur2014probing}, we do not explicitly define a scattering matrix here; instead, we analyze the scattered electric field to show that its change is frozen when a dipole's polarizability is changed.

We consider an ensemble of $D$ dipoles. The $i$th dipole is characterized by its polarizability $\alpha_i$. The application of an electric field $e_i$ at the location of the $i$th dipole induces a dipole moment $p_i=\alpha_i e_i$ in the $i$th dipole. The $D$ dipoles are coupled to each other via background Green's functions. $G_{ji}p_i$ is the field at the position of the $j$th dipole due to a dipole moment $p_i$ at the position of the $i$th dipole. If the background medium in which the $D$ dipoles are embedded is free space, $G_{ji}$ can be evaluated in closed form, and $G_{ji}=G_{ij}$ due to reciprocity. More generally, however, $G_{ji}$ can describe any arbitrary complex linear, static, passive background medium. If $\alpha_i$ is the radiation-reaction-corrected polarizability, $G_{ii}=0$ in free space whereas generally $G_{ii}\neq 0$ for a non-trivial background medium.

The superposition principle yields a set of coupled equations, $e_i = \alpha_i^{-1} p_i =  e_i^\mathrm{inc} + \sum_{j=1}^D G_{ij} p_j$, where $e_i^\mathrm{inc}$ is the incident field. We can cast this system of equations into matrix notation: $\mathbf{A}\mathbf{p} = \mathbf{e}^\mathrm{inc} + \mathbf{G}\mathbf{p}$. Here, we use $\mathbf{A}=\mathrm{diag}([\alpha_1^{-1},\dots,\alpha_D^{-1}])\in\mathbb{C}^{D\times D}$. Solving for $\mathbf{p}$ yields $\mathbf{p} = \mathbf{W}^{-1}\mathbf{e}^\mathrm{inc}$, where $\mathbf{W}=\mathbf{A}-\mathbf{G}$, such that $\mathbf{e}=(\mathbf{I}_D+\mathbf{G} \mathbf{W}^{-1})\mathbf{e}^\mathrm{inc}$. 
Changing the polarizability of the $i$th dipole alters the $i$th diagonal entry of $\mathbf{A}$, and thus also the $i$th entry of $\mathbf{W}$: $\hat{\mathbf{W}} = \mathbf{W}+(\hat{\alpha}_i^{-1}-{\alpha}_i^{-1})\mathbf{c}_i\mathbf{c}_i^\top$, where $\mathbf{c}_i$ is the $i$th canonical vector. According to the Sherman-Morrison identity~\cite{hager1989updating}, $\Delta\mathbf{e}= - \lambda \mathbf{f} \mathbf{g}^\top \mathbf{e}^\mathrm{inc}$, where $ \mathbf{f} = \mathbf{G}\mathbf{W}^{-1} \mathbf{c}_i$, $\mathbf{g}^\top = \mathbf{c}_i^\top \mathbf{W}^{-1}$ and $\lambda = (\hat{\alpha}_i^{-1}-{\alpha}_i^{-1}) / (1 + (\hat{\alpha}_i^{-1}-{\alpha}_i^{-1}) \mathbf{c}_i^\top \mathbf{W}^{-1}\mathbf{c}_i)$. Given $\Delta\mathbf{e}= - \lambda \mathbf{f} \mathbf{g}^\top \mathbf{e}^\mathrm{inc}$ and  recognizing that $\mathbf{g}^\top \mathbf{e}^\mathrm{inc}$ is a complex-valued scalar, it follows that the change of the scattered field is frozen because it is always collinear with $\mathbf{f}$, no matter the incident field $\mathbf{e}^\mathrm{inc}$.

\subsection*{Frozen differential transmission in MPLCs}

Let us consider for concreteness a common MPLC model~\cite{zhang2023multi}, describing the MPLC as scalar, forward-propagating 
diffraction through a cascade of $l$ infinitesimally thin, perfectly transmissive, phase-only masks separated by free-space propagation. The incoming wavefront $\mathbf{x}\in\mathbb{C}^{N_\mathrm{T}}$ is mapped to the outgoing wavefront $\mathbf{y}\in\mathbb{C}^{N_\mathrm{R}}$ by the linear operator $\mathbf{U}=\mathbf{P}_l\mathbf{M}_l\dots\mathbf{P}_1\mathbf{M}_1\mathbf{P}_0\in\mathbb{C}^{N_\mathrm{R}\times N_\mathrm{T}}$, where $\mathbf{M}_i=\mathrm{diag}([\mathrm{e}^{\jmath \phi_{i,1}},\dots,\mathrm{e}^{\jmath \phi_{i,M}}])\in\mathbb{C}^{M \times M}$ is the $i$th phase mask (all entries have unit magnitude) discretized into $M$ phase pixels, and $\mathbf{P}_i$ is the free-space propagation from the $i$th to the $(i+1)$th phase mask for $1\leq i<l$; $\mathbf{P}_0$ is the propagation from input to the first phase mask; $\mathbf{P}_l$ is the propagation from the last phase mask to the output. We emphasize that this model assumes the absence of any multiple scattering, be it within a phase mask or between phase masks. 

A large MPLC behaves like a random scattering medium~\cite{boucher2021full}. Changing the $j$th entry of the $i$th phase mask from $\mathrm{e}^{\jmath \phi_{i,j}}$ to $\mathrm{e}^{\jmath \hat{\phi}_{i,j}}$ results in a rank-one update of $\mathbf{U}$:
$\Delta\mathbf{U} = \mathbf{A}_i (\Delta\mathbf{M}_i) \mathbf{B}_i$, where $\mathbf{A}_i=\mathbf P_l\mathbf M_l\!\cdots\!\mathbf P_{i}\in\mathbb{C}^{N_\mathrm{R} \times M}$ and $\mathbf{B}_i=\mathbf P_{i-1}\mathbf M_{i-1}\!\cdots\!\mathbf P_1\mathbf M_1 \mathbf{P}_0\in\mathbb{C}^{M \times N_\mathrm{T}}$. Because all entries of $\Delta\mathbf{M}_i$ are zero except for the $j$th diagonal entry, $\Delta\mathbf{M}_i=(\mathrm{e}^{\jmath \hat{\phi}_{i,j}}-\mathrm{e}^{\jmath \phi_{i,j}})\mathbf{c}_j\mathbf{c}_j^\top$, where $\mathbf{c}_j$ is the $j$th canonical vector. 
Defining the column vector $\mathbf{f}_{i,j}=\mathbf{A}_i\mathbf{c}_j$ and the row vector $\mathbf{g}_{i,j}^\top=\mathbf{c}_j^\top\mathbf{B}_i$ yields $\Delta\mathbf{U} = (\mathrm{e}^{\jmath \hat{\phi}_{i,j}}-\mathrm{e}^{\jmath \phi_{i,j}}) \mathbf{f}_{i,j}  \mathbf{g}_{i,j}^\top$, clearly revealing the rank-one nature of $\Delta\mathbf{U}$. It follows as in our main text that $\Delta\mathbf{y} = (\Delta\mathbf{U}) \mathbf{x}$ is always collinear with $\mathbf{f}_{i,j}$, i.e., the differential output shape is frozen.

In contrast to the multiport-network model used in our main text, the MPLC model does not assume that the perturbation is point-like. Indeed, a phase mask pixel may be \emph{much larger} than the wavelength. The relevant constraint is \emph{sub-resolution}, not sub-wavelength: the modified region on the $i$th phase mask plane should fit within one joint diffraction-limited resolution cell set by the \emph{overlap} of the upstream and downstream angular acceptances. Detailed discussions of sampling requirements can be found in Ref.~\cite{zhang2023multi}.

\setcounter{figure}{0}
\renewcommand{\thefigure}{S\arabic{figure}}

\begin{figure*}
\centering
\includegraphics[width=\linewidth]{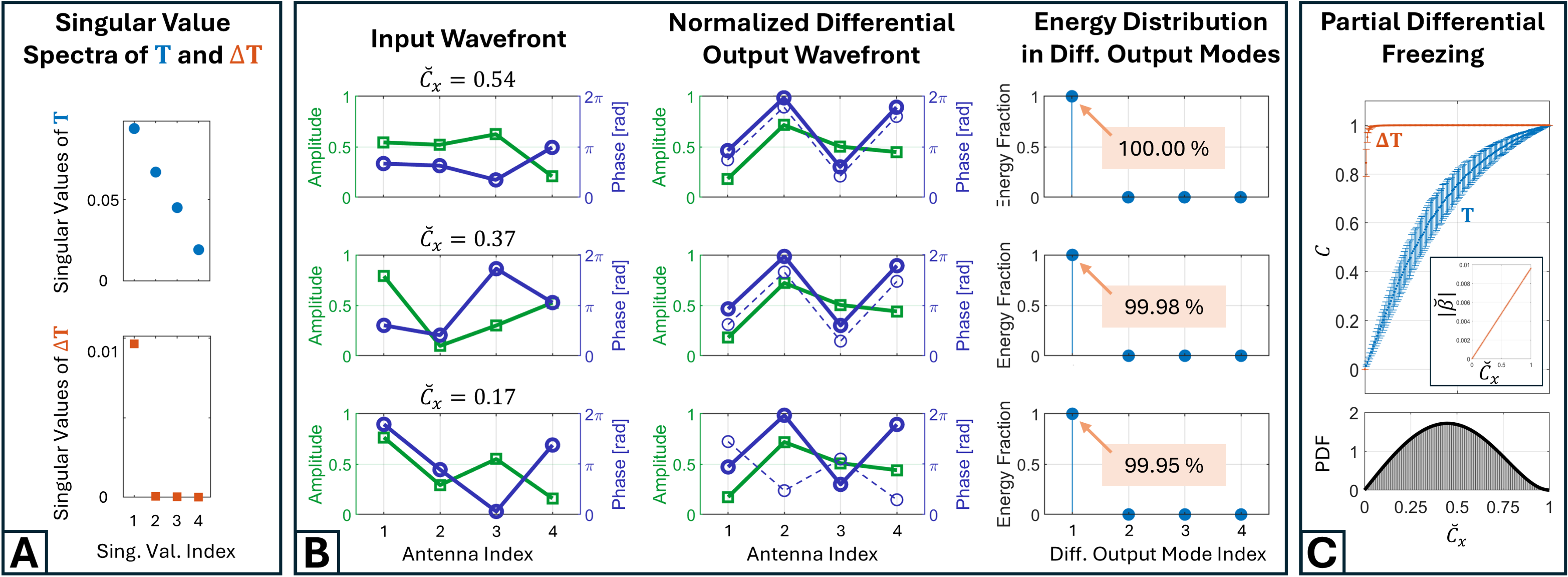}
\caption{Experimental observation of frozen differential transmission.
(A) Measured singular value spectra of $\mathbf{T}$ (top) and $\Delta\mathbf{T}$ (bottom).
(B) For three distinct input wavefronts (one per row), we display three items. 
Left column: $\mathbf{x}_\mathcal{T}$ in terms of amplitude (green, left axis) and phase (blue, right axis); $\breve{C}_x = | \mathbf{x}_\mathcal{T}^\dagger \breve{\mathbf{{r}}}_1 | / \lVert \mathbf{x}_\mathcal{T} \rVert_2 \lVert \breve{\mathbf{{r}}}_1 \rVert_2  $ indicates the overlap of $\mathbf{x}_\mathcal{T}$ with $\breve{\mathbf{r}}_1$.
Middle column: $\Delta\mathbf{y}_\mathcal{R}/\|\Delta\mathbf{y}_\mathcal{R}\|_2$ in terms of normalized amplitude (green, left axis), aligned phase (blue, right axis), and raw phase (blue dashed, right axis). The alignment consists in applying a global phase rotation $\mathrm{exp}(-\jmath\ \mathrm{arg}(\breve{\mathbf{s}}_1^\dagger \Delta\mathbf{y}_\mathcal{R}))$ to all entries of $\Delta\mathbf{y}_\mathcal{R}$.
Right column: Energy distribution of the normalized $\Delta\mathbf{y}_\mathcal{R}$ over the output modes (left singular vectors) of $\Delta\mathbf{T}$.
(C) Freezing metric $C$ as a function of $\breve{C}_x$ for $\Delta\mathbf{T}$ (red), and analogous results for $\mathbf{T}$ (blue). The inset shows the scaling of $|\breve{\beta}|$ with $\breve{C}_x$. The probability density function (PDF) of $\breve{C}_x$ for random wavefronts is also shown.
}
\label{FigS1}
\end{figure*}


\begin{thebibliography}{10}
\urlstyle{rm}
\expandafter\ifx\csname url\endcsname\relax
  \def\url#1{\texttt{#1}}\fi
\expandafter\ifx\csname urlprefix\endcsname\relax\def\urlprefix{URL }\fi
\expandafter\ifx\csname doiprefix\endcsname\relax\def\doiprefix{DOI: }\fi
\providecommand{\bibinfo}[2]{#2}
\providecommand{\eprint}[2][]{\url{#2}}

\bibitem{moustakas2000communication}
\bibinfo{author}{Moustakas, A.~L.}, \bibinfo{author}{Baranger, H.~U.},
  \bibinfo{author}{Balents, L.}, \bibinfo{author}{Sengupta, A.~M.} \&
  \bibinfo{author}{Simon, S.~H.}
\newblock \bibinfo{journal}{\bibinfo{title}{Communication through a diffusive
  medium: Coherence and capacity}}.
\newblock {\emph{\JournalTitle{Science}}} \textbf{\bibinfo{volume}{287}},
  \bibinfo{pages}{287--290} (\bibinfo{year}{2000}).

\bibitem{cao2022shaping}
\bibinfo{author}{Cao, H.}, \bibinfo{author}{Mosk, A.~P.} \&
  \bibinfo{author}{Rotter, S.}
\newblock \bibinfo{journal}{\bibinfo{title}{Shaping the propagation of light in
  complex media}}.
\newblock {\emph{\JournalTitle{Nat. Phys.}}} \textbf{\bibinfo{volume}{18}},
  \bibinfo{pages}{994--1007} (\bibinfo{year}{2022}).

\bibitem{shi2012transmission}
\bibinfo{author}{Shi, Z.} \& \bibinfo{author}{Genack, A.~Z.}
\newblock \bibinfo{journal}{\bibinfo{title}{Transmission eigenvalues and the
  bare conductance in the crossover to {Anderson} localization}}.
\newblock {\emph{\JournalTitle{Phys. Rev. Lett.}}}
  \textbf{\bibinfo{volume}{108}}, \bibinfo{pages}{043901}
  (\bibinfo{year}{2012}).

\bibitem{davy2012focusing}
\bibinfo{author}{Davy, M.}, \bibinfo{author}{Shi, Z.} \&
  \bibinfo{author}{Genack, A.~Z.}
\newblock \bibinfo{journal}{\bibinfo{title}{Focusing through random media:
  Eigenchannel participation number and intensity correlation}}.
\newblock {\emph{\JournalTitle{Phys. Rev. B}}} \textbf{\bibinfo{volume}{85}},
  \bibinfo{pages}{035105} (\bibinfo{year}{2012}).

\bibitem{wang2011transport}
\bibinfo{author}{Wang, J.} \& \bibinfo{author}{Genack, A.~Z.}
\newblock \bibinfo{journal}{\bibinfo{title}{Transport through modes in random
  media}}.
\newblock {\emph{\JournalTitle{Nature}}} \textbf{\bibinfo{volume}{471}},
  \bibinfo{pages}{345--348} (\bibinfo{year}{2011}).

\bibitem{pena2014single}
\bibinfo{author}{Pe{\~n}a, A.}, \bibinfo{author}{Girschik, A.},
  \bibinfo{author}{Libisch, F.}, \bibinfo{author}{Rotter, S.} \&
  \bibinfo{author}{Chabanov, A.}
\newblock \bibinfo{journal}{\bibinfo{title}{The single-channel regime of
  transport through random media}}.
\newblock {\emph{\JournalTitle{Nat. Commun.}}} \textbf{\bibinfo{volume}{5}},
  \bibinfo{pages}{3488} (\bibinfo{year}{2014}).

\bibitem{leseur2014probing}
\bibinfo{author}{Leseur, O.}, \bibinfo{author}{Pierrat, R.},
  \bibinfo{author}{S{\'a}enz, J.} \& \bibinfo{author}{Carminati, R.}
\newblock \bibinfo{journal}{\bibinfo{title}{Probing two-dimensional {Anderson}
  localization without statistics}}.
\newblock {\emph{\JournalTitle{Phys. Rev. A}}} \textbf{\bibinfo{volume}{90}},
  \bibinfo{pages}{053827} (\bibinfo{year}{2014}).

\bibitem{chizhik2000capacities}
\bibinfo{author}{Chizhik, D.}, \bibinfo{author}{Foschini, G.~J.} \&
  \bibinfo{author}{Valenzuela, R.~A.}
\newblock \bibinfo{journal}{\bibinfo{title}{Capacities of multi-element
  transmit and receive antennas: Correlations and keyholes}}.
\newblock {\emph{\JournalTitle{Electron. Lett.}}}
  \textbf{\bibinfo{volume}{36}}, \bibinfo{pages}{1099--1100}
  (\bibinfo{year}{2000}).

\bibitem{chizhik2002keyholes}
\bibinfo{author}{Chizhik, D.}, \bibinfo{author}{Foschini, G.~J.},
  \bibinfo{author}{Gans, M.~J.} \& \bibinfo{author}{Valenzuela, R.~A.}
\newblock \bibinfo{journal}{\bibinfo{title}{Keyholes, correlations, and
  capacities of multielement transmit and receive antennas}}.
\newblock {\emph{\JournalTitle{IEEE Trans. Wirel. Commun.}}}
  \textbf{\bibinfo{volume}{1}}, \bibinfo{pages}{361--368}
  (\bibinfo{year}{2002}).

\bibitem{almers2003measurement}
\bibinfo{author}{Almers, P.}, \bibinfo{author}{Tufvesson, F.} \&
  \bibinfo{author}{Molisch, A.~F.}
\newblock \bibinfo{journal}{\bibinfo{title}{Measurement of keyhole effect in a
  wireless multiple-input multiple-output ({MIMO}) channel}}.
\newblock {\emph{\JournalTitle{IEEE Commun. Lett.}}}
  \textbf{\bibinfo{volume}{7}}, \bibinfo{pages}{373--375}
  (\bibinfo{year}{2003}).

\bibitem{almers2006keyhole}
\bibinfo{author}{Almers, P.}, \bibinfo{author}{Tufvesson, F.} \&
  \bibinfo{author}{Molisch, A.~F.}
\newblock \bibinfo{journal}{\bibinfo{title}{Keyhole effect in {MIMO} wireless
  channels: Measurements and theory}}.
\newblock {\emph{\JournalTitle{IEEE Trans. Wirel. Commun.}}}
  \textbf{\bibinfo{volume}{5}}, \bibinfo{pages}{3596--3604}
  (\bibinfo{year}{2006}).

\bibitem{salemeh2025invariance}
\bibinfo{author}{Salemeh, E.}, \bibinfo{author}{F{\'e}lix, S.} \&
  \bibinfo{author}{Pagneux, V.}
\newblock \bibinfo{journal}{\bibinfo{title}{Invariance of the speckle pattern
  of the transmitted wave in periodic waveguides}}.
\newblock {\emph{\JournalTitle{Sci. Rep.}}} \textbf{\bibinfo{volume}{15}},
  \bibinfo{pages}{2504} (\bibinfo{year}{2025}).

\bibitem{salemeh2025freezing}
\bibinfo{author}{Salemeh, E.}, \bibinfo{author}{F{\'e}lix, S.} \&
  \bibinfo{author}{Pagneux, V.}
\newblock \bibinfo{journal}{\bibinfo{title}{Freezing of the transmitted wave
  pattern through gratings}}.
\newblock {\emph{\JournalTitle{Wave Motion}}} \bibinfo{pages}{103606}
  (\bibinfo{year}{2025}).

\bibitem{morizur2010programmable}
\bibinfo{author}{Morizur, J.-F.} \emph{et~al.}
\newblock \bibinfo{journal}{\bibinfo{title}{Programmable unitary spatial mode
  manipulation}}.
\newblock {\emph{\JournalTitle{J. Opt. Soc. Am. A}}}
  \textbf{\bibinfo{volume}{27}}, \bibinfo{pages}{2524--2531}
  (\bibinfo{year}{2010}).

\bibitem{berkhout2010efficient}
\bibinfo{author}{Berkhout, G.~C.}, \bibinfo{author}{Lavery, M.~P.},
  \bibinfo{author}{Courtial, J.}, \bibinfo{author}{Beijersbergen, M.~W.} \&
  \bibinfo{author}{Padgett, M.~J.}
\newblock \bibinfo{journal}{\bibinfo{title}{Efficient sorting of orbital
  angular momentum states of light}}.
\newblock {\emph{\JournalTitle{Phys. Rev. Lett.}}}
  \textbf{\bibinfo{volume}{105}}, \bibinfo{pages}{153601}
  (\bibinfo{year}{2010}).

\bibitem{zhang2023multi}
\bibinfo{author}{Zhang, Y.} \& \bibinfo{author}{Fontaine, N.~K.}
\newblock \bibinfo{journal}{\bibinfo{title}{Multi-plane light conversion: a
  practical tutorial}}.
\newblock {\emph{\JournalTitle{arXiv:2304.11323}}}  (\bibinfo{year}{2023}).

\bibitem{zhou2021large}
\bibinfo{author}{Zhou, T.} \emph{et~al.}
\newblock \bibinfo{journal}{\bibinfo{title}{Large-scale neuromorphic
  optoelectronic computing with a reconfigurable diffractive processing unit}}.
\newblock {\emph{\JournalTitle{Nat. Photonics}}} \textbf{\bibinfo{volume}{15}},
  \bibinfo{pages}{367--373} (\bibinfo{year}{2021}).

\bibitem{boucher2021full}
\bibinfo{author}{Boucher, P.}, \bibinfo{author}{Goetschy, A.},
  \bibinfo{author}{Sorelli, G.}, \bibinfo{author}{Walschaers, M.} \&
  \bibinfo{author}{Treps, N.}
\newblock \bibinfo{journal}{\bibinfo{title}{Full characterization of the
  transmission properties of a multi-plane light converter}}.
\newblock {\emph{\JournalTitle{Phys. Rev. Res.}}} \textbf{\bibinfo{volume}{3}},
  \bibinfo{pages}{023226} (\bibinfo{year}{2021}).

\bibitem{wyatt1968differential}
\bibinfo{author}{Wyatt, P.~J.}
\newblock \bibinfo{journal}{\bibinfo{title}{Differential light scattering: a
  physical method for identifying living bacterial cells}}.
\newblock {\emph{\JournalTitle{Appl. Opt.}}} \textbf{\bibinfo{volume}{7}},
  \bibinfo{pages}{1879--1896} (\bibinfo{year}{1968}).

\bibitem{wyatt1969identification}
\bibinfo{author}{Wyatt, P.~J.}
\newblock \bibinfo{journal}{\bibinfo{title}{Identification of bacteria by
  differential light scattering}}.
\newblock {\emph{\JournalTitle{Nature}}} \textbf{\bibinfo{volume}{221}},
  \bibinfo{pages}{1257--1258} (\bibinfo{year}{1969}).

\bibitem{bustamante1980circular}
\bibinfo{author}{Bustamante, C.}, \bibinfo{author}{Maestre, M.~F.} \&
  \bibinfo{author}{Tinoco~Jr, I.}
\newblock \bibinfo{journal}{\bibinfo{title}{Circular intensity differential
  scattering of light by helical structures. {I. T}heory}}.
\newblock {\emph{\JournalTitle{J. Chem. Phys.}}} \textbf{\bibinfo{volume}{73}},
  \bibinfo{pages}{4273--4281} (\bibinfo{year}{1980}).

\bibitem{bustamante1980circularII}
\bibinfo{author}{Bustamante, C.}, \bibinfo{author}{Maestre, M.~F.} \&
  \bibinfo{author}{Tinoco~Jr, I.}
\newblock \bibinfo{journal}{\bibinfo{title}{Circular intensity differential
  scattering of light by helical structures. {II. A}pplications}}.
\newblock {\emph{\JournalTitle{J. Chem. Phys.}}} \textbf{\bibinfo{volume}{73}},
  \bibinfo{pages}{6046--6055} (\bibinfo{year}{1980}).

\bibitem{bustamante1983circular}
\bibinfo{author}{Bustamante, C.}, \bibinfo{author}{Tinoco~Jr, I.} \&
  \bibinfo{author}{Maestre, M.~F.}
\newblock \bibinfo{journal}{\bibinfo{title}{Circular differential scattering
  can be an important part of the circular dichroism of macromolecules.}}
\newblock {\emph{\JournalTitle{Proc. Natl. Acad. Sci. U.S.A.}}}
  \textbf{\bibinfo{volume}{80}}, \bibinfo{pages}{3568--3572}
  (\bibinfo{year}{1983}).

\bibitem{forker2012optical}
\bibinfo{author}{Forker, R.}, \bibinfo{author}{Gruenewald, M.} \&
  \bibinfo{author}{Fritz, T.}
\newblock \bibinfo{journal}{\bibinfo{title}{Optical differential reflectance
  spectroscopy on thin molecular films}}.
\newblock {\emph{\JournalTitle{Annu. Rep. Prog. Chem., Sect. C: Phys. Chem.}}}
  \textbf{\bibinfo{volume}{108}}, \bibinfo{pages}{34--68}
  (\bibinfo{year}{2012}).

\bibitem{7163589}
\bibinfo{author}{Cha, M.}, \bibinfo{author}{Phillips, R.~D.},
  \bibinfo{author}{Wolfe, P.~J.} \& \bibinfo{author}{Richmond, C.~D.}
\newblock \bibinfo{journal}{\bibinfo{title}{Two-stage change detection for
  synthetic aperture radar}}.
\newblock {\emph{\JournalTitle{IEEE Trans. Geosci. Remote Sens.}}}
  \textbf{\bibinfo{volume}{53}}, \bibinfo{pages}{6547--6560}
  (\bibinfo{year}{2015}).

\bibitem{wahl2016new}
\bibinfo{author}{Wahl, D.~E.}, \bibinfo{author}{Yocky, D.~A.},
  \bibinfo{author}{Jakowatz, C.~V.} \& \bibinfo{author}{Simonson, K.~M.}
\newblock \bibinfo{journal}{\bibinfo{title}{A new maximum-likelihood change
  estimator for two-pass {SAR} coherent change detection}}.
\newblock {\emph{\JournalTitle{IEEE Trans. Geosci. Remote Sens.}}}
  \textbf{\bibinfo{volume}{54}}, \bibinfo{pages}{2460--2469}
  (\bibinfo{year}{2016}).

\bibitem{8401705}
\bibinfo{author}{Monti-Guarnieri, A.~V.} \emph{et~al.}
\newblock \bibinfo{journal}{\bibinfo{title}{Coherent change detection for
  multipass {SAR}}}.
\newblock {\emph{\JournalTitle{IEEE Trans. Geosci. Remote Sens.}}}
  \textbf{\bibinfo{volume}{56}}, \bibinfo{pages}{6811--6822}
  (\bibinfo{year}{2018}).

\bibitem{alouini_dpsk}
\bibinfo{author}{Simon, M.} \& \bibinfo{author}{Alouini, M.-S.}
\newblock \bibinfo{journal}{\bibinfo{title}{A unified approach to the
  probability of error for noncoherent and differentially coherent modulations
  over generalized fading channels}}.
\newblock {\emph{\JournalTitle{IEEE Trans. Commun.}}}
  \textbf{\bibinfo{volume}{46}}, \bibinfo{pages}{1625--1638}
  (\bibinfo{year}{1998}).

\bibitem{miller2019waves}
\bibinfo{author}{Miller, D. A.~B.}
\newblock \bibinfo{journal}{\bibinfo{title}{Waves, modes, communications, and
  optics: a tutorial}}.
\newblock {\emph{\JournalTitle{Adv. Opt. Photonics}}}
  \textbf{\bibinfo{volume}{11}}, \bibinfo{pages}{679--825}
  (\bibinfo{year}{2019}).

\bibitem{roy2007effective}
\bibinfo{author}{Roy, O.} \& \bibinfo{author}{Vetterli, M.}
\newblock \bibinfo{journal}{\bibinfo{title}{The effective rank: A measure of
  effective dimensionality}}.
\newblock {\emph{\JournalTitle{Proc. EUSIPCO}}} \bibinfo{pages}{606--610}
  (\bibinfo{year}{2007}).

\bibitem{del2025ambiguity}
\bibinfo{author}{del Hougne, P.}
\newblock \bibinfo{journal}{\bibinfo{title}{Ambiguity-aware segmented
  estimation of mutual coupling in large {RIS}: Algorithm and experimental
  validation}}.
\newblock {\emph{\JournalTitle{arXiv:2507.22750}}}  (\bibinfo{year}{2025}).

\bibitem{sol2024experimentally}
\bibinfo{author}{Sol, J.}, \bibinfo{author}{Prod’Homme, H.},
  \bibinfo{author}{Le~Magoarou, L.} \& \bibinfo{author}{del Hougne, P.}
\newblock \bibinfo{journal}{\bibinfo{title}{Experimentally realized
  physical-model-based frugal wave control in metasurface-programmable complex
  media}}.
\newblock {\emph{\JournalTitle{Nat. Commun.}}} \textbf{\bibinfo{volume}{15}},
  \bibinfo{pages}{2841} (\bibinfo{year}{2024}).

\bibitem{del2025experimental}
\bibinfo{author}{del Hougne, P.}
\newblock \bibinfo{journal}{\bibinfo{title}{Experimental multiport-network
  parameter estimation and optimization for multi-bit {RIS}}}.
\newblock {\emph{\JournalTitle{arXiv:2507.02168}}}  (\bibinfo{year}{2025}).

\bibitem{prod2023efficient}
\bibinfo{author}{Prod’homme, H.} \& \bibinfo{author}{del Hougne, P.}
\newblock \bibinfo{journal}{\bibinfo{title}{Efficient computation of
  physics-compliant channel realizations for (rich-scattering)
  {RIS}-parametrized radio environments}}.
\newblock {\emph{\JournalTitle{IEEE Commun. Lett.}}}
  \textbf{\bibinfo{volume}{27}}, \bibinfo{pages}{3375--3379}
  (\bibinfo{year}{2023}).

\bibitem{sol2025optimal}
\bibinfo{author}{Sol, J.}, \bibinfo{author}{Le~Magoarou, L.} \&
  \bibinfo{author}{del Hougne, P.}
\newblock \bibinfo{journal}{\bibinfo{title}{Optimal blind focusing on
  perturbation-inducing targets in sub-unitary complex media}}.
\newblock {\emph{\JournalTitle{Laser Photonics Rev.}}}
  \textbf{\bibinfo{volume}{19}}, \bibinfo{pages}{2400619}
  (\bibinfo{year}{2025}).

\bibitem{del2025virtual}
\bibinfo{author}{del Hougne, P.}
\newblock \bibinfo{journal}{\bibinfo{title}{Virtual {VNA}: Minimal-ambiguity
  scattering matrix estimation with a fixed set of “virtual” load-tunable
  ports}}.
\newblock {\emph{\JournalTitle{IEEE Trans. Instrum. Meas.}}}
  \textbf{\bibinfo{volume}{74}}, \bibinfo{pages}{1--19} (\bibinfo{year}{2025}).

\bibitem{ahmed2024over}
\bibinfo{author}{Ahmed, I.}, \bibinfo{author}{Davy, M.},
  \bibinfo{author}{Prod’homme, H.}, \bibinfo{author}{Besnier, P.} \&
  \bibinfo{author}{del Hougne, P.}
\newblock \bibinfo{journal}{\bibinfo{title}{Over-the-air emulation of
  electronically adjustable {Rician} {MIMO} channels in a
  programmable-metasurface-stirred reverberation chamber}}.
\newblock {\emph{\JournalTitle{IEEE Trans. Antennas Propag.}}}
  \textbf{\bibinfo{volume}{73}}, \bibinfo{pages}{2104--2119}
  (\bibinfo{year}{2025}).

\bibitem{wolf1982new}
\bibinfo{author}{Wolf, E.}
\newblock \bibinfo{journal}{\bibinfo{title}{New theory of partial coherence in
  the space--frequency domain. {Part I}: {S}pectra and cross spectra of
  steady-state sources}}.
\newblock {\emph{\JournalTitle{J. Opt. Soc. Am. A}}}
  \textbf{\bibinfo{volume}{72}}, \bibinfo{pages}{343--351}
  (\bibinfo{year}{1982}).

\bibitem{wolf1986new}
\bibinfo{author}{Wolf, E.}
\newblock \bibinfo{journal}{\bibinfo{title}{New theory of partial coherence in
  the space-frequency domain. {Part II}: {S}teady-state fields and higher-order
  correlations}}.
\newblock {\emph{\JournalTitle{J. Opt. Soc. Am. A}}}
  \textbf{\bibinfo{volume}{3}}, \bibinfo{pages}{76--85} (\bibinfo{year}{1986}).

\bibitem{withington2002modal}
\bibinfo{author}{Withington, S.} \& \bibinfo{author}{Murphy, J.~A.}
\newblock \bibinfo{journal}{\bibinfo{title}{Modal analysis of partially
  coherent submillimeter-wave quasi-optical systems}}.
\newblock {\emph{\JournalTitle{IEEE Trans. Antennas Propag.}}}
  \textbf{\bibinfo{volume}{46}}, \bibinfo{pages}{1651--1659}
  (\bibinfo{year}{2002}).

\bibitem{zhang2019scattering}
\bibinfo{author}{Zhang, H.}, \bibinfo{author}{Hsu, C.~W.} \&
  \bibinfo{author}{Miller, O.~D.}
\newblock \bibinfo{journal}{\bibinfo{title}{Scattering concentration bounds:
  brightness theorems for waves}}.
\newblock {\emph{\JournalTitle{Optica}}} \textbf{\bibinfo{volume}{6}},
  \bibinfo{pages}{1321--1327} (\bibinfo{year}{2019}).

\bibitem{roques2024measuring}
\bibinfo{author}{Roques-Carmes, C.}, \bibinfo{author}{Fan, S.} \&
  \bibinfo{author}{Miller, D.~A.}
\newblock \bibinfo{journal}{\bibinfo{title}{Measuring, processing, and
  generating partially coherent light with self-configuring optics}}.
\newblock {\emph{\JournalTitle{Light Sci. Appl.}}}
  \textbf{\bibinfo{volume}{13}}, \bibinfo{pages}{260} (\bibinfo{year}{2024}).

\bibitem{guo2024unitary}
\bibinfo{author}{Guo, C.} \& \bibinfo{author}{Fan, S.}
\newblock \bibinfo{journal}{\bibinfo{title}{Unitary control of partially
  coherent waves. {I.} {A}bsorption}}.
\newblock {\emph{\JournalTitle{Phys. Rev. B}}} \textbf{\bibinfo{volume}{110}},
  \bibinfo{pages}{035430} (\bibinfo{year}{2024}).

\bibitem{guo2024unitary2}
\bibinfo{author}{Guo, C.} \& \bibinfo{author}{Fan, S.}
\newblock \bibinfo{journal}{\bibinfo{title}{Unitary control of partially
  coherent waves. {II.} {T}ransmission or reflection}}.
\newblock {\emph{\JournalTitle{Phys. Rev. B}}} \textbf{\bibinfo{volume}{110}},
  \bibinfo{pages}{035431} (\bibinfo{year}{2024}).

\bibitem{harling2025optical}
\bibinfo{author}{Harling, M.}, \bibinfo{author}{Stevenson, C.},
  \bibinfo{author}{Toussaint, K.~C.} \& \bibinfo{author}{Abouraddy, A.~F.}
\newblock \bibinfo{journal}{\bibinfo{title}{Optical communications through
  highly scattering channels using the coherence-rank}}.
\newblock {\emph{\JournalTitle{APL Photonics}}} \textbf{\bibinfo{volume}{10}}
  (\bibinfo{year}{2025}).

\bibitem{twiss1955nyquist}
\bibinfo{author}{Twiss, R.}
\newblock \bibinfo{journal}{\bibinfo{title}{Nyquist's and {T}hevenin's theorems
  generalized for nonreciprocal linear networks}}.
\newblock {\emph{\JournalTitle{J. Appl. Phys.}}} \textbf{\bibinfo{volume}{26}},
  \bibinfo{pages}{599--602} (\bibinfo{year}{1955}).

\bibitem{bosma1967theory}
\bibinfo{author}{Bosma, H.}
\newblock \bibinfo{journal}{\bibinfo{title}{On the theory of linear noisy
  systems}}.
\newblock {\emph{\JournalTitle{Technische Hogeschool Eindhoven}}}
  (\bibinfo{year}{1967}).

\bibitem{haus2012electromagnetic}
\bibinfo{author}{Haus, H.~A.}
\newblock \emph{\bibinfo{title}{Electromagnetic noise and quantum optical
  measurements}} (\bibinfo{publisher}{Springer Science \& Business Media},
  \bibinfo{year}{2012}).

\bibitem{wedge2002noise}
\bibinfo{author}{Wedge, S.~W.} \& \bibinfo{author}{Rutledge, D.~B.}
\newblock \bibinfo{journal}{\bibinfo{title}{Noise waves and passive linear
  multiports}}.
\newblock {\emph{\JournalTitle{IEEE Microw. Guid. Wave Lett.}}}
  \textbf{\bibinfo{volume}{1}}, \bibinfo{pages}{117--119}
  (\bibinfo{year}{1991}).

\bibitem{miller2017universal}
\bibinfo{author}{Miller, D. A.~B.}, \bibinfo{author}{Zhu, L.} \&
  \bibinfo{author}{Fan, S.}
\newblock \bibinfo{journal}{\bibinfo{title}{Universal modal radiation laws for
  all thermal emitters}}.
\newblock {\emph{\JournalTitle{Proc. Natl. Acad. Sci. U.S.A.}}}
  \textbf{\bibinfo{volume}{114}}, \bibinfo{pages}{4336--4341}
  (\bibinfo{year}{2017}).

\bibitem{sweeney2020theory}
\bibinfo{author}{Sweeney, W.~R.}, \bibinfo{author}{Hsu, C.~W.} \&
  \bibinfo{author}{Stone, A.~D.}
\newblock \bibinfo{journal}{\bibinfo{title}{Theory of reflectionless scattering
  modes}}.
\newblock {\emph{\JournalTitle{Phys. Rev. A}}} \textbf{\bibinfo{volume}{102}},
  \bibinfo{pages}{063511} (\bibinfo{year}{2020}).

\bibitem{ona2024orthogonal}
\bibinfo{author}{O{\~n}a, D.}, \bibinfo{author}{Ortega-Gomez, A.},
  \bibinfo{author}{Hern{\'a}ndez, O.} \& \bibinfo{author}{Liberal, I.}
\newblock \bibinfo{journal}{\bibinfo{title}{Orthogonal thermal noise and
  transmission signals: A new coherent perfect absorption’s feature}}.
\newblock {\emph{\JournalTitle{Phys. Rev. Lett.}}}
  \textbf{\bibinfo{volume}{133}}, \bibinfo{pages}{103801}
  (\bibinfo{year}{2024}).

\bibitem{hammami2025statistical}
\bibinfo{author}{Hammami, C.}, \bibinfo{author}{Le~Magoarou, L.} \&
  \bibinfo{author}{del Hougne, P.}
\newblock \bibinfo{journal}{\bibinfo{title}{Statistical multiport-network
  modeling and efficient discrete optimization of {RIS}}}.
\newblock {\emph{\JournalTitle{arXiv:2508.01776}}}  (\bibinfo{year}{2025}).

\bibitem{lax1952multiple}
\bibinfo{author}{Lax, M.}
\newblock \bibinfo{journal}{\bibinfo{title}{Multiple scattering of waves. II.
  The effective field in dense systems}}.
\newblock {\emph{\JournalTitle{Phys. Rev.}}} \textbf{\bibinfo{volume}{85}},
  \bibinfo{pages}{621} (\bibinfo{year}{1952}).

\bibitem{hager1989updating}
\bibinfo{author}{Hager, W.~W.}
\newblock \bibinfo{journal}{\bibinfo{title}{Updating the inverse of a matrix}}.
\newblock {\emph{\JournalTitle{SIAM Rev.}}} \textbf{\bibinfo{volume}{31}},
  \bibinfo{pages}{221--239} (\bibinfo{year}{1989}).

\end{thebibliography}
\end{document}